\documentclass{article}
\usepackage[cp1251]{inputenc}\usepackage[russian]{babel}
\usepackage[dvips]{graphics}
\usepackage{array,hhline,amssymb,amsthm,color,amsfonts}
\def\*{*\!\!}
\def\L{{\mathfrak{L}}}
\def\O{{\mathcal{O}}}  \def\o{{\mathbf{o}}}
\def\K{{\overline{\O}}}  \def\k{{\overline{\o}}}
   
\def\R{\mathbb{R}}  \def\C{\mathbb{C}}    
\def\d12{\raise0.4ex\hbox{\scriptsize 1}\!/\!\lower0.4ex\hbox{\scriptsize 2}}
\def\tr{{\rm tr}} 
\def\bux#1#2{\buildrel{*}\over{#1}_{#2}}
\title{Induction of gravity and strong interactions on non-associative algebra}
\date{}
\author{V. Yu. Dorofeev\thanks{Russia, Saint-Petersburg, e-mail: friedlab@mail.ru}\\
Friedmann laboratory for Theoretical physics}
\begin{document}
\maketitle
\begin{abstract}
In this work we take view on space-time as dual representation of fields on manifold. Given we accept such view, the space of functions in operator representation becomes probability amplitudes f(x) of a particle. Since the probabilistic interpretation of f(x) module follows from f(x) and f*(x) duality, the problem of finding representation of wave function reduces to Frobenius theorem for division algebras. We then construct the gravity model and the model of strong interactions as a logical implication of the extension of fields interaction algebra to non-associative algebra of octonions. The application of Penrouse approach in spinor representation of space-time combined with the proposed approach further leads us to chirality of interactions and absence of right neutrino and left spinors in singleton state. We also show the obtained model is consistent with GRG and SU(3) for distances larger than Planck length. The obtained model also contains the generalization of GRG to the model with torsion given the chirality of the field.
Importantly, the application of our approach gives us Einstein equation with gravity constant equal to squared Planck mass as consequence of non-associativity of interactions fields.
We construct strong interaction model based on matrix representation of octonion algebra with special multiplication rule. The strong interaction is induced by vector fields of octonion algebra with constant gravity force, which is in turn induced by a pair of charged $W$ vector bosons from standard theory of weak interactions. The problem of quark confinement is presented as a spinor solution inside ``black hole''.
\end{abstract}

\section*{Введение}
Современные теории Великого объединения строятся на переносе
известных и хорошо зарекомендовавших себя методов квантовой теории
поля на единую модель гравитации, электрослабых и сильных
взаимодействий. Например, теория суперсимметрии строится на основе
алгебраического подхода над градуированной алгеброй \cite{Vess}, при
этом пространством являются как скалярные так и спинорные
переменные. Но обязательно ли определено пространство переменных?
Например, имеется известный скачок описания от классической физики к
квантовой физике, при этом пока нет гладкого перехода между этими
двумя описаниями природы. Автор попытался в этой работе найти общие
принципы того, какие же должны быть переменные. Результатом
рассуждений в этом направлении явилось построение модели, основанной
на алгебре октонионов. Вообще же, обобщением алгебры кватернионов,
имеющим операцию деления, является алгебра октонионов, по
терминологии Артура Кэли, или алгебра октав, по терминологии Джона
Грейвса. Известно \cite{Baez}, что первая публикация, описывающая
эту алгебру, принадлежит А. Кэли и датирована 1845 годом -- через
два года после открытия кватернионов Гамильтоном.

Видимо первая попытка применения алгебры октонионов к физической
теории была сделана в работе \cite{Jordan}, где вводилась
октонионная квантовая механика. Однако серьёзного развития идеи этой
работы не имели, возможно из-за недостаточного развития
математических и физический концепций тех лет, а возможно в
математической абстрактности используемого в статье алгебраического
аппарата октонионов. Немного ранее, в работах Цорна \cite{Zorn},
были предложены матричные представления алгебры октонионов, но
элементами матриц Цорна были специальные объекты: числа и векторы.
Поэтому эта работа указывала на путь представления неассоциативной
алгебры матрицами, но предлагаемый вид матриц был очень абстрактным.
Не замеченной физиками, по мнению автора, явилась замечательная
статья J. Debout и R. Delbago \cite{DD}, которая буквально
<<вписывала>> октонионы в физическую теорию: было найдено
представление октав матрицами Дирака с особым правилом умножения.

Актуальность проведённых автором исследований расширения групповой
теории Вайн\-берга-Салама слабых взаимодействий на неассоциативную
алгебру обусловлена необходимостью поиска новых подходов для
построения теории, совмещающих в себе как общекоординатные, так и
калибровочные преобразования. При этом имеется надежда включить
группу общекоординатных преобразований как проявление
неассоциативности взаимодействия.

Основные  успехи в теории элементарных частиц связаны с
использованием теории представлений групп. В данной работе модель
взаимодействий строится не на базе заложенных симметрий, а
посредством расширения алгебраической структуры поля. С этой целью
как естественное следствие некоторых рассуждений в Стандартной
Теории (СТ) Вайнберга-Салама предлагается заменить   симметрию
$SU(2)\times U(1)$ на неассоциативную алгебру \cite{Dor1}. В силу
неассоциативности  алгебры происходит потеря свойств калибровочной
симметрии, которая, будем считать, обусловлена структурой вакуума на
планковском уровне, вне которых они восстанавливаются.
Дополнительные эффекты, которые возникают как следствие
неассоциативности, предлагается интерпретировать, в частности,
проявлением гравитации.

Замечательным  обстоятельством предлагаемого подхода является
органичное вхождение в развиваемую теорию группы цветной симметрии
$SU(3)$ как группы на $G_2$. Другое интересное обстоятельство -- то,
что кварки строятся как решение под горизонтом Шварцшильда с
двумерной внутренней топологией типа струны, также органично
вписываемую в данный подход. Это позволяет объяснить отсутствие
свободных кварков как решения внутри <<чёрной дыры>>, что вполне
согласуется с теоремой Хоукинга о невозможности извлечения
какой-либо информации из-под горизонта, кроме как информации о
массе, заряде и моменте \cite{Hawking}.

\section{Волновая функция частицы как функция на многообразии}
Определим существование некоторого  векторного поля $A^d$
аксиоматически. (В СТ Вайнберга-Салама вводится поле Хиггса
$\varphi(x)$. Тогда вектор $\partial_\mu\varphi(x)$ может играть
роль этого векторного поля. Однако в приводимых ниже рассуждениях
конкретный вид векторного поля не предполагается.) Размерность
линейного пространства $T$, в котором определено поле $A^d$
определим как размерность физического 4-х мерного
пространства-времени и считаем, что пространство-время -- линейное
пространство, определённое полем $A^d$. Таким образом физическое
пространство-время -- это пространст\-венно-временное многобразие
$\Omega$ с соответствующей топологией. (Под линейным
пространством-временем понимаем самоидентификацию поля $A^d$.
Физическое пространство-время -- арена проявления поля $A^d$ и
возможно других физических полей.)

Возможность описывать движение  в пространственно-временном
многообразии $\Omega$ рассматриваем как метки поля $A^d$. Могут
существовать и другие поля $A$, которые в пространстве-времени
проявляются как векторные. Пусть физическое поле $A^d$ воздействует
по крайней мере само на себя, индуцируя <<точечные сгущения>> $x$ на
$\Omega\ni x$, которые позволяют различать соседние точки
многообразия $\Omega$ (<<точечными сгущениями>> называем результат
самодействия поля $A^d$ и его взаимодействие с полями $A$). Так как
<<точечные сгущения>> индуцируют метки, то относительно них с
помощью линейной формы будем строить дуальное векторное пространство
$T^*(x)$. (Если нет <<точечных сгущений>>, то нет и дуального
пространства). Векторное поле $A^d$ на многообразии $\Omega$
понимается обычным образом: существует однопараметрическая функция
$\tilde\varphi(x(\tau))$ на физическом многообразии $\Omega$,
относительно которой задаётся векторное поле как $d/d\tau$. Тогда в
дуальном векторном пространстве определяем ортогональный базис
линейных форм $dx=*A^d$ из правила $<dx,A^d>=I$, где $<dx,A^d>$ --
линейная форма с единичной матрицей $I$. Здесь форма
$dx=(dx_0,\dots,dx_3)$ индуцирует координаты $(\xi_0,\dots,\xi_3)$.
(Физически речь идёт о координатах $(t,x,y,z)=(\xi_0,\dots,\xi_3)$.)
Формально строим кинематические соотношения между координатами вида
\begin{equation}\label{p1_01}
dx_i/dx_0=v_i,(i=1,2,3)
\end{equation}
и экспериментально устанавливаем динамические  уравнения в случае
консервативных систем вида
\begin{equation}\label{p1_02}
mdv_i/dx_0=\delta_{ij}\partial_j V^{ext}(x),(i=1,2,3,I=(\delta_{ij}))
\end{equation}
(здесь $V^{ext}(x)$ -- внешнее потенциальное  поле, например,
ньютоновское гравитационное или кулоновское электрическое)
соответствующие ньютоновской механике и обобщаем всё на специальную
теорию относительности, вводя лоренцевую инвариантность интервала
\begin{equation}\label{p1_03}
ds^2=dt^2-dx^2-dy^2-dz^2
\end{equation}

С точке зрения предлагаемого подхода в (\ref{p1_02}) речь идёт о
некотором токе $J$ как о векторе, который определяется
идентификацией <<точек сгущения>>. В базисе $e_\mu,\mu=0,1,2,3$ он
имеет вид:
\begin{equation}\label{p1_04}
J=J_\mu e^\mu=m\frac{dx_\mu}{ds}e^\mu
\end{equation}

<<Замечая>>, что <<точечные сгущения>> обладают зарядом, строим
напряжённость электромагнитного поля $F$, как антисимметричный
тензор второй степени. Из опыта убеждаемся в точности формы $F$, то
есть $dF=0$. Равенство $dF=0$ понимаем  как отсутствие источников
поля, с одной стороны, с другой стороны, это означает существование
1-формы $\omega$, для которой $d\omega=F$. В частном случае
электромагнитного поля тензор первой степени $\omega$ -- это
электромагнитный вектор-потенциал $A^{el-m}=\omega$ и
$dA^{el-m}=F^{el-m}$. Будем рассматривать электромагнитный
вектор-потенциал $A^{el-m}$ и векторное поле $A^d$ как два
линейно-независимых векторных поля линейного пространства $T$. (Пока
считаем, что пространственно-временное многообразие $\Omega$
совпадает с линейным пространством $T$.) Так как вектор, дуальный
$dx$ -- это вектор $\partial/\partial x=\partial_x$, то векторами
пространства $T$ в некотором представлении являются векторы
$\partial_x$ и $A$, которые взаимодействуют с <<точками сгущения>>
-- состояниями, обозначаемыми в дальнейшем как $\Phi$ (или $\Psi$).
Тогда можно говорить о взаимодействии полей $A^d$ и $A$ с
<<точечными сгущениями>> $\Phi$ в виде $A\cdot\Phi$ и
$A^d\cdot\Phi$. Вектор, дуальный вектору $A^d$ -- это вектор $dx$.
Вектор, дуальный вектору $dx$, это вектор $\partial_x$. Таким
образом можно установить изоморфизм между $\partial_x$ и $A^d$.
Вектор $\partial_x$ задаётся на пространстве функций, поэтому будем
считать, что в выбранном нами представлении
$A^d\cdot\Phi=\partial_x\Phi$.

Состояние $\Phi$ является формальным представлением <<точечных
сгущений>> -- состояния частицы. Следовательно определено дуальное
пространство состояний $*\Phi$, для которого $<*\Phi,\Phi>=C$.
(Можно договориться, что $C=1$.) Например, в пространстве функций
равенство $<*\Phi,\Phi>=1$ могло бы выглядеть как
$\int(*\Phi(x))\Phi(x)dx=1$.

Пусть речь идёт об одной частице в объёме $V$. Тогда состояние
$\Phi$ не должно зависеть от координаты $x$, если по-прежнему
понимать наличие аргумента $x$ в выражении $\Phi(x)$ как указатель
на локализацию частицы в точке с координатой $x$. В этом случае
получаем <<хороший>> (с точки зрения предъявляемых на данный момент
требований к функциям $\Phi(x)$ и $*\Phi(x)$) вид этих функций:
$\Phi(x)=*\Phi(x)=1/\sqrt V$. Фактически имеем равенство:
\begin{equation}\label{p1_1}
<*\Phi,\Phi>=\int(*\Phi(x))\Phi(x)dx=\int\Phi(x)\Phi(x)dx=\int(*\Phi(x))*\Phi(x)dx=1
\end{equation}

Данное равенство замечательно тем, что предлагает способ
симметризации состояний частиц и состояний античастиц.
\begin{equation}\label{p1_2}
|\Phi\cdot*\Phi|=|\Phi|\cdot|*\Phi|=1
\end{equation}

(\ref{p1_2}) и (\ref{p1_1}) позволяют интерпретировать величину
$|\Phi(x)|^2$ как вероятность нахождения частицы в состоянии $\Phi$
в точке $x$, а само значение (\ref{p1_2}) как полную вероятность. В
теории вероятностей математическое ожидание непрерывной случайной
величины $\hat\xi$, имеющий закон распределения $\tilde
F_{\tilde\xi}$, когда $\tilde F_{\tilde\xi}'(t)=
\rho_{\tilde\xi}(t)$, находится по формуле
\begin{equation}\label{p1_7}
M(\hat\xi)=\int\hat\xi(x)\rho_\xi(x)dx,
\end{equation}
поэтому все характеристики частиц находятся как среднии по ансамблю.
При этом частицы рассматриваются как безструктурные. В
статистической физике такой подход реализуется в операторном
формализме физических величин над вероятностным распределением
частиц, например, в \cite{Kuni}.

Если  под состоянием $\Phi$ понимать пространство состояний,
соответствующее частицам, то с современных позиций логично назвать
пространство состояний $*\Phi$ состоянием античастиц. Действительно,
если говорить, что <<точечные сгущения>> порождают частицы,
обладающие некоторыми свойствами, то сохранение этих свойств требует
состояния античастиц, которые описываются <<одновременно>> с
состоянием $\Phi$.

Состояние $\Phi$ инвариантно  относительно ряда преобразований (по
крайней мере относительно лоренцевых преобразований) так как под ним
понимается описание, например, частицы в обычном
пространстве-времени. Введём физическое поле $*F$ в дуальном
пространстве. Для него $d*F=4\pi J$, где $J$ -- ток, индуцированный
<<точечными сгущениями>>. Появление тока $J$ обусловлено появлением
<<точек сгущения>>, то есть состояний $\Phi$, поэтому теперь
$d*F\ne0$. Вектор тока $J$ определим как
$J=(d*\Phi)\Phi-*\Phi(d\Phi)$. Такое определение согласовано с тем,
что $dd*F=d^2*F=0$. Таким образом имеется следующая совокупность
утверждений:
\begin{equation}\label{p1_3}
dF=0
\end{equation}
\begin{equation}\label{p1_4}
d*F=4\pi J
\end{equation}
\begin{equation}\label{p1_5}
J=(d*\Phi)\Phi-*\Phi(d\Phi)
\end{equation}
\begin{equation}\label{p1_6}
*\Phi\cdot\Phi=1
\end{equation}

Рассмотрим локализованное состояние частицы  в объёме $V$. Например,
$\Phi(x)=C_0e^{-kx}$. Логично понимать свободное движение частицы и
античастицы как состояния, которые описываются одинаковым образом.
Следовательно состояния частиц и античастиц описываются одной и той
же функцией. Тогда $*\Phi(x)=\Phi(x)$. Такой подход не является
конструктивным так как возвращает к уже изученному движению частиц в
статистической физике, когда они считаются безструктурными.
Следовательно необходимо изменить структуру функции состояния
$\Phi(x)$. Но имеется и конструктивный подход, сохраняющий симметрию
состояний частиц и античастиц, но расширяющий и углубляющий эту
симметрию: рассмотреть различные алгебры, для которых выполняется
равенство (\ref{p1_2}), то есть сформулировать

{\it{\bf Гипотеза состояния частицы.}  Алгебра представления
состояния частиц $a$ и $b$, с нормами $|a|$ и $|b|$ соответственно,
удовлетворяет равенству}
\begin{equation}\label{p1_8}
|a\cdot b|=|a|\cdot|b|
\end{equation}

По Теореме Гурвица: {\it Любая  нормированная алгебра с единицей (то
есть выполняется условие (\ref{p1_8})) изоморфна одной из четырёх
алгебр: действительных чисел, комплексных чисел, кватернионов или
октав.}

В предлагаемом подходе естественно  понимать статистическую механику
как вещественное представление пространства состояний, так как в
основном речь едёт о механических движениях, не учитывающих
существование различных свойств у частиц и античастиц.

Из (\ref{p1_6}) следует, что состояние  вида $\Phi(x)=Ce^{ikx}$,
удовлетворяет <<Гипотезе представления состояния частицы>>. В этом
случае величиной, дуальной времени $dt$ является
$\displaystyle\frac\hbar i\partial_t$. Заметим естественное введение
мнимой постоянной $i$. Понятно, что постоянная Планка вводится из
более глубоких соображений -- здесь достаточно понимание того, что
дуальный элемент $\partial_t$ определяется с точностью до
постоянного множителя.

Предложенное обобщение на комплексные  числа является единственным
минимальным обощением по теореме Гурвица.

Будем считать определённой линейную  комбинацию векторных полей
$A^d$ и $A$ вида $A^d\cdot e_1+A\cdot e_2$ в некотором изоморфном
представлении ($q$ -- заряд частицы)
\begin{equation}\label{pv_9}
\frac\hbar i\partial_x+qA(x)
\end{equation}

Симметрия $U(1)$ для нормы $\Psi^*(x)\Psi(x)$  индуцирует локальную
симметрию $U(1)$ для выражения
\begin{equation}\label{pv_10}
\Psi^*(x)(\frac\hbar i\partial_x+qA(x))\Psi(x)
\end{equation}
при этом выражение $\displaystyle\frac\hbar i\partial_x+qA(x)$ в
смысле геометродинамики эквивалентно ковариантной производной в
главном расслоении.

Первоначально электромагнитный вектор-потенциал $A^{el-m}$ был
определён из точности тензора напряжённости поля $F^{el-m}$. Теперь,
однако, понятно, что первичной характеристикой является всё-таки
вектор-потенциал $A^{el-m}$. Тем самым тензор $F^{el-m}$ необходимо
определить как производную от $A$. С этой целью заметим, что в
пространстве Минковского имеется билинейная квадратичная формы с
метрикой Минковского, поэтому может быть определена площадка
$\Sigma=dx\wedge dx$, при обходе которой ковариантным образом
($\displaystyle\frac\hbar i\nabla\Psi=A\Psi$) индуцируется величина
\cite{Fadeev}
\begin{equation}\label{p1_11}
\Delta\Psi(x)=F\Sigma\Psi
\end{equation}

Задача совмесного состояния тензорного поля $F$  и ему  сопряжённого
$*F$ по определению является согласованной, поэтому может быть
определён скаляр $L=*F\wedge F$. Таким образом получаем ещё
несколько скалярных инвариантов
\begin{equation}\label{p1_12}
S_{el.-m.}=*F\wedge F
\end{equation}
\begin{equation}\label{p1_13}
S_{sp}=(\nabla*\Psi)\Psi-*\Psi(\nabla\Psi)
\end{equation}
\begin{equation}\label{p1_14}
S_{sc}=*(\nabla\Phi)\nabla\Phi
\end{equation}

Действие электромагнитного поля (\ref{p1_12}) имеет вид
\begin{equation}\label{p1_15}
S_{el.-m.}=*F\wedge F=\varepsilon^{pqjl}F_{pq}dx_j\wedge dx_l\wedge F_{ik}dx^i\wedge dx^k=$$
$$=\eta^{ij}\eta^{kl}F_{ik}F_{jl}d\Omega=\eta^{ij}\eta^{kl}(A_{k,i}^0-A_{i,k}^0)\Sigma^0\cdot(A_{l,j}^0-A_{j,l}^0)\Sigma^0d\Omega=$$
$$=\eta^{ij}\eta^{kl}(A_{k,i}^0-A_{i,k}^0)(A_{l,j}^0-A_{j,l}^0)\Sigma^0d\Omega
\end{equation}
(Здесь введено представление  вещественных чисел $\Sigma^0=1$.
Дуальное пространство в (\ref{p1_12}) строится как изоморфизм в
пространствах со скалярным произведением ко-пространств и дуальных
пространств. Наконец возникающий коэффициент после суммирования
внесён с элемент объёма $d\Omega$.)

Тут же возникает замечание:  <<точки сгущения>>, с которыми
ассоциируются состояния $\Psi$, оказываются вторичным элементом по
отношению к векторному полю. Векторное поле $A^d$ оказывает
воздействие на состояния $\Psi$. Первоначальные рассуждения были
нужны, чтобы понять требования на состояния $\Psi$. Естественным
является такое же требование на векторные поля $A$ как элементов той
же алгебры. Поэтому <<алгебра векторных полей $A$ должна
принадлежать одной из алгебр: вещественной, комплексной,
кватернионной или алгебре октав>>. В нашем случае какому-либо из
представлений этих алгебр. Это утверждение назовём <<Гипотезой об
алгебре поля>>. Но векторное поле $A$ является <<первичным>>
объектом. Следовательно <<Гипотеза о состоянии частицы>> возможно
неверна, но нарушать <<Гипотезу об алгебре поля>> оснований нет.

Вектор поля $A^d$ не имеет заранее  определённой структуры. Его
проявление определяется самодействием и взаимодействием с полями
$A$. Результат взаимодействия с математической точки зрения является
элемент алгебры. Далее нашей задачей было построение
непротиворечивой схемы над этой алгеброй и изучение её свойств. По
теореме Гурвица можно сделать следующий шаг, не противоречащий
общему подходу, излагаемому в этой работе. Считать, что алгебра --
это алгебра кватернионов. Тогда элементом пространства состояний
является алгебраический элемент, изоморфный алгебре кватернионов.
Формально в качестве такого элемента можно было бы взять любое
представление, но ввиду того, что уже <<придумано>> хорошо
согласующееся с опытом описание модели над комплексными функциями (в
соответствии с предлагаемыми рассуждениями комплексные величины
описываю внутреннюю степень свободы, идентифицирующуюся как заряд
для представления частиц и античастиц) будем требовать, чтобы в
качестве коммутативного поля над функциями состояния выступало
комплексное поле $C$. По этой причине выберем в качестве такого
представление представление матрицами $SU(2)$. Тогда вектор поля $A$
должен быть элементом этой алгебры. Следовательно, будем считать,
что
\begin{equation}\label{p1_16}
A=A^{\mu(a)}(x)\cdot\sigma^ae_\mu
\end{equation}
$\sigma^a$ -- это единичная матрица  и три матрицы Паули.
Коэффициенты $A^{\mu(a)}_\mu(x)$ выбираем вещественными, так как с
ними ассоциируем физические величины.

Следовательно пространство состояний  является элементом
пространства, на котором действует группа $SU(2)$. Но тогда форма
$dx$ имеет вид:
\begin{equation}\label{p1_17}
dx=dx_\mu^a\cdot\sigma^ae^\mu
\end{equation}

Фактически мы имеем спинорное представление  пространства-времени
вида \cite{Penrose}
\begin{equation}\label{p1_18}
\left(\matrix{T+Z&X+iY\cr X-iY&T-Z}\right)=T\sigma^0+X\sigma^1+Y\sigma^2+Z\sigma^3
\end{equation}

В \cite{Penrose} показано, что изоморфное  отображение спиноров в
группу вращений $SO(3)$ можно установить, вводя изотропный флаг $\bf
K$ и пространственно-подобный вектор $\bf L$ (в обозначениях
\cite{Penrose}). Тогда плоскость $\bf\Pi$, натянутая на векторы $\bf
K$ и $\bf L$:
\begin{equation}\label{p1_19}
{\bf\Pi}=a{\bf K}+b{\bf L},\quad b>0
\end{equation}
отвечает лоренц-преобразованиям и бустам  пространства-времени
\begin{equation}\label{p1_20}
V=Te_t+Xe_x+Ye_y+Ze_z
\end{equation}

Группа преобразований $SU(2)$ имеет односвязную компоненту  и
двухсвязную компоненту, отвечающие фундаментальной группе $\pi_1$ и
$\pi_2$. Соответствующее представление строится из спиноров и
антиспиноров (в обозначениях \cite{Penrose} элементы $\sigma^{AA'}$,
где индексы $A,A'$ принимают значения 0,1 и соответствуют
комплексно-сопряжённым базисным элементам). В нашем случае это
означает введение биспиноров в качестве пространства состояний. С
другой стороны, ограничение $b>0$ можно учесть если заметить, что
вектор $\bf L$ отвечает пространственно-подобному вектору. На
единичной сфере $T=1$ он имеет координаты ${\bf L}=(0,x,y,z)$ и при
пространственных отражениях переходит в $-\bf L$, что необходимо
запретить, так как $b>0$ в (\ref{p1_19}). Так как наше синглетное
представление (соответствующее фундаментальной группе $\pi_1$) --
это волновая функция частиц, то дублетное представление -- волновая
функция античастиц (для сохранения заряда). Если учесть условие
$b>0$, вводя правые биспиноры для синглетных состояний, то левые
биспиноры античастиц будут соответствовать дублетам. Актуален вопрос
почему именно левое нейтрино и правое синглетное состояние? Играет
роль спиновое взаимодействие (\ref{p12_71}). Дело в том, что
энергетически более выгодно иметь направление спина по
радиус-вектору, который простанственно-подобен и направлен в
будущее. Тогда дублету необходимо приписать левый знак спиральности.
Нейтрино же будучи безмассовой частицей фиксирует общую
спиральность. Заметим, что к выбору направления спина у дублета
формула (\ref{p12_71}) или ей аналогичная не оказывает воздействия,
так как общий спин пары антинейтрино и электрона (нейтрино и
позитрона) может быть сделан равным нулю.

Действия (\ref{p1_12}-\ref{p1_14}), обобщённые на кватернионную
структуру поля и состояния, когда поля (\ref{pv_9}) берутся в
представлении трёх матриц Паули и единичной матрицы $2\times2$ в
киральном случае позволяет построить модель электрослабых
взаимодействий. Правда для перенормируемости теории с массивными
частицами в полное действие необходимо включить самодействие
состояний $\Phi$ в виде потенциала Хиггса
\begin{equation}\label{p1_21}
V_{Higgs}=m^2|\Phi|^2-\frac f4|\Phi|^4
\end{equation}
что в общем-то не нарушает общий взгляд на  построение теории,
излагаемый здесь. В результате мы приходит к киральной модели слабых
взаимодействий $SU(1)\times SU(2)$ теории Вайнберга-Салама.

С точки зрения предлагаемого здесь подхода,  в соответствии с
теоремой Гурвица, остался один, последний шаг -- обобщение теории на
неассоциативную алгебру октонионов.

\begin{equation}\label{p1_24}
A=A^{\mu(a)}(x)\cdot\Sigma^ae_\mu
\end{equation}
$\Sigma^a$ -- это некоторые четыре матрицы  представления алгебры
октонионов. Коэффициенты $A^{\mu(a)}(x)$ выбираем вещественными, так
как с ними ассоциируем физические величины.

Алгебра октонионов может быть  реализована с помощью симметрии
$G_2$, которую можно редуцировать до симметрии $SU(3)$. Симметрия
$SU(3)$ имеет три независимые подалгебры $SU(2)$, что можно
рассматривать как три поколения лептонных и кварковых полей.
Следовательно пространство состояний является элементом
пространства, на котором действует группа $SU(2)$. Но тогда форма
$dx$ имеет вид (\ref{p1_17}).

Но возникает серьёзная проблема неассоциативности  поля. Как
следствие оказывается невозможным извлечение симметрии из каждого из
выражений действий (\ref{p1_12}-\ref{p1_14}). Только действие
(\ref{p1_13}) может быть представлено в ассоциативной форме. На заре
развития теории электромагнетизма основными понятиями
электромагнитной теории были измеримые поля $\vec E$ и $\vec B$.
Если пойти тем же путём и считать первичным именно эти поля, то
действие (\ref{p1_12}) оказывается квадратичным по физическим полям
$\bf E$ и $\bf B$:
\begin{equation}\label{p1_22}
F=dA+[A,A]=F_{ik}dx_i\wedge dx_k=$$ $$((A_{k,i}^a-A_{i,k}^a)\Sigma^a+[A_{k,i}^a\Sigma^a,A_{i,k}^b\Sigma^b])dx_i\wedge dx_k=$$ $$=((A_{k,i}^a-A_{i,k}^a)\Sigma^a+A_{k,i}^aA_{i,k}^b[\Sigma^a,\Sigma^b])dx_i\wedge dx_k=$$
$$=((A_{k,i}^a-A_{i,k}^a)\Sigma^a+A_{k,i}^aA_{i,k}^b[\Sigma^a,\Sigma^b])dx_i\wedge dx_k=$$
$$=((A_{k,i}^a-A_{i,k}^a)\Sigma^a+A_{k,i}^aA_{i,k}^b\varepsilon^{abc}\Sigma^c)dx_i\wedge dx_k=F^a_{ik}\Sigma^adx_i\wedge dx_k
\end{equation}
-- использовано обозначение образующих алгебры  $\Sigma^a$ со
структурными постоянными $\varepsilon^{abc}$. Неассоциативность
определяется на трёх элементах алгебры, для которых в общем случае:
\begin{equation}\label{p1_23}
a\cdot(b\cdot c)\ne(a\cdot b)\cdot c
\end{equation}
но (\ref{p1_22})  линейно по физическому полю $F^a$, поэтому
квадратичен лагранжиан (\ref{p1_12}). С другой стороны, алгебра
октонионов со своими структурными постоянными образует групповую
симметрию $G_2$ и содержит симметрию $SU(3)$. В этом смысле
октонионная структура поля в (\ref{p1_12}) должна проявляться в том
числе и как структура, содержащая физическую симметрию $SU(3)$
\cite{India}. Но симметрия $SU(3)$ наблюдается как цветная симметрия
в теории сильных взаимодействий. По-видимому исследования,
аналогичные сделанным в \cite{Penrose} позволили бы пролить свет
свет на ряд удивительных фактов в теории сильных взаимодействий, но
пока такая работа не проводилась и здесь этот аспект не исследуется.

Опыт обобщения теории на алгебру кватернионов показывает,  что
конструктивным является представление именно матрицами $(2\times2)$.
Октонионы тоже могут быть представлены матрицами. Данное
представление рассмотрено в следующем параграфе. Заметим, что
алгебраическое обобщение функции состояния частиц в силу приведённых
рассуждений требует обобщения представления пространства-времени на
неассоциативную алгебру, редукция которого на пространство-время
Минковского (\ref{p1_20}) приводит к ограничению на возможные
состояния частиц (по аналогии с появлением P-неинвариантности в
(\ref{p1_19}), требующим фиксированной спиральности мультиплета). Но
этот вопрос выходит за пределы данной работы.

Естественным следствием приведённой выше логики построения теории
является тот факт, что некоторого общего уравнения поля не должно
быть. Имеются свободные поля $A$ общей алгебры (некая алгебра, более
широкая чем алгебра октав Кэлли). Результатом взаимодействия полей
являются состояния материи, для которых в рамках выбранной алгебры
определяются инвариантные структуры, соответствующие лагранжиану.
Попытка искать ответ в увеличении числа симметрий может быть
конструктивной только для определённых структур (как в теории
твёрдого тела дополнительные симметрии индуцируют дополнительные
свойства лагранжиана и его модельный вид).

Приведённые  выше рассуждения указывают на некоторую математическую
логику современного представления квантовой механики. Автор не
ставит задачу построения жёсткой схемы, но пытается вписать своё
обобщение современной теории поля таким образом, чтобы можно было
говорить о преемственности его точки зрения, излагаемой здесь, и
современных представлений квантовой теории поля.

В работе  не рассматривается квантовый вторично-квантованный аспект
предлагаемого подхода. Дело в том, что представление частиц вполне
соответствует духу излагаемого материала и означает разве что
альтернативный выбор пространства состояний частиц. Проблема же
расходимостей, которая могла бы волновать, в данном подходе видимо
отсутствует (см. параграф о снятии неассоциативности), так как в
силу неассоциативности октонионов позволяет изменять знак слагаемых
бесконечного ряда нужным образом. Наверняка это не слабость теории,
а её сила. Тем не менее работа в этом направлении должна быть
проведена аккуратно.

Ранее указывалось \cite{Dor5}, что предлагаемый здесь подход  удобно
назвать О-теорией.

\section{Матричное представление октонионов}
Октонионы или октавы Кэли образуют алгебру \cite{Baez}.  Эта алгебра
может быть получена методом удвоения Кэли-Диксона из алгебры
кватернионов и обычно представляется как гиперкомплексное число в
восьмимерном линейном пространстве над полем вещественных чисел.
Имеется матричное представление октонионов \cite{Zorn}, но с особым
правилом умножения. В дальнейшем мы будем использовать именно
матричное представление октав Кэли в виде матриц Дирака,
предложенное в работе \cite{DD}. Такое пространство представления
обозначим как $\tilde\O$. Тогда
\begin{equation}\label{p2_1}
\tilde\O=\{\forall\o=\sum_{A=0}^7\alpha^A\tilde\Sigma^A=\alpha^A\tilde\Sigma^A,\quad\alpha^A\in\R\}
\end{equation}
где $\tilde\Sigma^K$ -- антиэрмитовы матрицы, которые
удовлетворяют правилу умножения октонионов ($I,J,K=1,2,\dots,7$):
\begin{equation}\label{p2_2}
\tilde\Sigma^I\cdot\tilde\Sigma^J=-\delta^{IJ}+\varepsilon^{IJK}\tilde\Sigma^K,
\end{equation}

Здесь введён полностью антисимметричный тензор $\varepsilon^{IJK}$,
у которого ненулевыми значениями являются только
$$\varepsilon^{123}=\varepsilon^{145}=\varepsilon^{176}=\varepsilon^{246}=
\varepsilon^{257}=\varepsilon^{347}=\varepsilon^{365}=1$$ и значения
$\varepsilon^{IJK}$, получаемые перестановкой  индексов. Матрица
$\tilde\Sigma^0$ -- обычная единичная матрица.

Заметим, что базисные элементы алгебры $\tilde\O$ неассоциативны:
$$\{\tilde\Sigma^A,\tilde\Sigma^B,\tilde\Sigma^d\}=(\tilde\Sigma^A\tilde\Sigma^B)\tilde\Sigma^d-
\tilde\Sigma^A(\tilde\Sigma^B\tilde\Sigma^d)=2\varepsilon^{ABCD}\tilde\Sigma^D$$
где полностью антисимметричный тензор $\varepsilon^{IJKL}$  отличен
от нуля только для следующих элементов:
\begin{equation}\label{p2_3}
\varepsilon^{1247}\;=\;\varepsilon^{1265}\;=\;\varepsilon^{2345}
\;=\;\varepsilon^{2376}\;=\;\varepsilon^{3146}\;=\;\varepsilon^{3157}
\;=\;\varepsilon^{4567}\;=1
\end{equation}

Вместо матриц представления мнимых единиц пространства  октонионов
удобно воспользоваться эрмитовыми матрицами
$\Sigma^K=i\tilde\Sigma^K,K=1,2,$ $\dots,7$. С точностью до мнимой
единицы структурные постоянные умножения таких матриц не изменяются
в сравнении со структурными константами умножения антиэрмитовых
матриц (\ref{p2_2}). При этом новое пространство является
комплексификацией исходного пространства. Обозначим это пространство
как $\O$. Всякий элемент этого пространства имеет вид:
\begin{equation}\label{p2_4}
\O=\{\forall\o=\alpha^A\Sigma^A,\quad\alpha^A\in\C,A=0,1,\dots,7\}
\end{equation}

Здесь введено восемь эрмитовых матриц ($I=1,2,3$)
\begin{equation}\label{p2_5}
\matrix{\Sigma^0=\left(\matrix{\sigma^0&0\cr0&\sigma^0}\right)&\Sigma^I=
\left(\matrix{0&-i\sigma^I\cr i\sigma^I&0}\right)\cr
\Sigma^4=\left(\matrix{-\sigma^0&0\cr0&\sigma^0}\right)&
\Sigma^{4+I}=\left(\matrix{0&-\sigma^I\cr -\sigma^I&0}\right)}
\end{equation}
где $\sigma^0$ -- единичная матрица и $\sigma^I,I=1,2,3$ -- матрицы
Паули:
\begin{equation}\label{p2_6}
\sigma^1=\left(\matrix{0&1\cr1&0}\right),\quad
\sigma^2=\left(\matrix{0&-i\cr i&0}\right),\quad
\sigma^3=\left(\matrix{1&0\cr0&-1}\right),\quad
\sigma^0=\left(\matrix{1&0\cr0&1}\right)
\end{equation}

Матрицы $\Sigma^A,A=0,1,2,\dots,7$ линейно независимы  и образуют
восьмимерное линейное пространство $\O$ над полем комплексных чисел.
На пространстве $\O$ введём операцию умножения по правилу,
следующему из (\ref{p2_2}). В результате мы получим алгебру $\O$.

Для двух произвольных элементов линейного пространства  $\O$
определим правило умножения:
$$\o*\o'=\left(\matrix{\lambda I&A\cr B&\xi I}\right)*
\left(\matrix{\lambda' I&A'\cr B'&\xi' I}\right)=$$
\begin{equation}\label{p2_7}
=\left(\matrix{(\lambda\lambda'+\frac12\tr(AB'))I\hfill&\lambda
A'+\xi' A+\frac i2[B,B']\hfill\cr\lambda'B+\xi B'-\frac
i2[A,A']&(\xi\xi'+\frac12\tr(BA'))I\hfill}\right)
\end{equation}

Здесь введены матрицы $A,A',B,B$ над комплексным полем чисел.
Размерности этих матриц $(2\times2)$. Введены также единичная
матрица $I=\sigma^0$ и комплексные числа
$\lambda,\xi,\lambda',\xi'$.

Правило умножения матриц (\ref{p2_7}) делает построенную  алгебру
матриц неассоциативной по умножению.

Введём ещё одно линейное пространство
\begin{equation}\label{p2_8}
\O_\R=\{\forall\o=\alpha^A\Sigma^A,\quad\alpha^A\in\R,A=0,1,\dots,7\}
\end{equation}

Пространство $\O_\R$ уже не образует алгебры так как  произведение
двух элементов из $\O_\R$ могут не принадлежать $\O_\R$.

Если базис линейного пространства $\O$ дополнить двумя матрицами
\begin{equation}\label{p2_9}
f^8=\left(\matrix{0&I\cr I&0}\right),\quad f^9=\left(\matrix{0& iI\cr-iI&0}\right)
\end{equation}
то мы получим расширенное пространство $\K$
\begin{equation}\label{p2_10}
\K=\{\forall\o=\sum\alpha^Af^A,\quad\alpha^A\in\C,A=0,1,\dots,9\}
\end{equation}
в котором определено произведение (\ref{p2_7}).

Определим пространство $\K$ -- как неассоциативный $\O$-модуль, где $\O$ -- кольцо.

С помощью умножения (\ref{p2_7}) на расширенном пространстве $\K$ введём свёртку:
\begin{equation}\label{p2_11}
(\k_1,\k_2)=\frac12\tr(\k_1^+*\k_2 )=\frac12\tr(\left(\matrix{\lambda_1^*I&B_1^+\cr A_1^+&\xi_1^*I}\right)*
\left(\matrix{\lambda_2I&A_2\cr B_2&\xi_2I}\right))$$
$$=\lambda_1^*\lambda_2+\xi_1^*\xi_2+\frac12\tr(B_1^+B_2))
+\frac12\tr(A_1^+A_2)
\end{equation}
отображающую $\K\times\K$ в $\C$. Заметим, что данная  свёртка
определяет скалярное произведение и норму на $\K$.

\section{Модели снятия неассоциативности октонионов}
Умножая  октонионы, необходимо следить за порядком умножения. Можно
предложить  схему умножения октонионов, придавая ей физические
основания. Конечно отдавая себе отчёт, что всякая схема является
частным случаем некоторого более общего результата. Тем не менее
представляет интерес рассмотреть разные схемы определения порядка
умножения октонионов и увидеть физические следствия, к которым можно
прийти.

Можно предложить, например, три модели снятия неассоциативности:

{\bf Вероятностная модель.}
\begin{equation}\label{p3_1}
prob(A*B*C*D)=p_1((A*B)*C)*D+p_2A*(B*C)*D
\end{equation}
$$p_1+p_2=1$$

Здесь введены частоты $p_1$ и $p_2$.  Их значения определяются
количеством перестановок скобок в (\ref{p3_1}), дающих одиниковый
результат. Например, в случае (\ref{p3_1}) логично считать, что
$p_1=3/4,p_2=1/4$ так как три варианта перестановки скобок дают
одинаковый ответ в первом члене правой части (\ref{p3_1}) и только
один вариант -- во втором члене правой части (\ref{p3_1}) (в
представлении октонионов матрицами).

Определение (\ref{p3_1}) распространяется и на большее число элементов.

{\bf Минимальная модель.}
\begin{equation}\label{p3_2}
min(A*B*C*D)=\min\{A*(B*C)*D,(A*B*C)*D\}
\end{equation}

Правая часть (\ref{p3_2}) означает выбор такой расстановки скобок,
при котором значение $A*B*C*D$ минимально.

Аналогично определяется {\bf максимальная модель.}

Выбор знака при снятии неассоциативности методом максимальной или
минимальной модели оставляет возможность выбирать знак по своему
ус\-мотрению, поэтому допустимо определить принцип {\bf минимакса}
как выбор по принципу минимальной или максимальной модели. Иногда
такой принцип выбора разумно определять как {\bf потенциальный}.
Насколько правильным окажется выбор для теоретической модели можно
судить только по результатам этого выбора.

Очевидно, что в случае ассоциативности мы получаем тот же результат,
что и в определённых выше моделях.

{\bf Смешанная модель.}
\begin{equation}\label{p3_3}
mix(A*B*C*D)=p\cdot min(A*B*C*D)+q\cdot prob(A*B*C*D)
\end{equation}
$$p+q=1$$

Смысл принципа снятия неассоциативности в этом методе ясен из определения.

\section{Лагранжиан О-теории}
Так как лагранжиан, основанный на обобщении  алгебры векторных полей
до кватернионов в киральном представлении спиноров $SU(2)\times
U(1)$, не описывает гравитационные и сильные взаимодействия, то
рассмотрим его обобщение на алгебру октонионов. Таким образом, пусть
поля возбуждения $A^d$ определены на алгебре октонионов. Тогда
дуальное ему пространство также строится на алгебре октонионов.
Следовательно координатное пространство как дуальное по аналогии с
(\ref{p1_03}) могут быть представлены в виде:
\begin{equation}\label{p4_01}
V=T\Sigma_t+X\Sigma_x+Y\Sigma_y+Z\Sigma_z
\end{equation}
Здесь $\Sigma_t,\Sigma_x,\Sigma_y,\Sigma_z$ --  элементы
представления алгебры октонионов. В частности,
$\Sigma^0,\Sigma^1,\Sigma^2,\Sigma^3$ из (\ref{p2_5}) образуют
подалгебру алгебры октонионов, изоморфную алгебре кватернионов.
Тогда ограничимся матрицами Паули и единичной матрицей для
представления координатного пространства. В данном случае мы опять
получаем киральное предсталение с левыми дублетами и правыми
синглетами.

Введём компенсирующие поля ${\bf A_a}(x),x\in M_4,a=0,1,2,3$  в
пространстве Минковского на пространстве матриц $\O_\R$:
\begin{equation}\label{p4_1}
{\bf A_a}(x)=A_a^A(x)\Sigma^A,A=0,1,2,\dots,7
\end{equation}
где $A_a^A(x)$ -- вещественные векторные поля.

В работе \cite{Dor1} предложен лагранжиан, обобщающий  лагранжиан
Вайн\-берга-Салама на альтернативное кольцо $\O$ с неассоциативным
модулем $\K$ в виде ($A,B=0,1,\dots,7,K=1,2,\dots,7$):
$$\L_\o=\L_f+(\partial_a\bux\Psi\varphi-
\frac i2q^AA_a^A\bux\Psi\varphi*\Sigma^A)
*(\partial^a\Psi_\varphi+
\frac i2q^BA^{a(B)} \Sigma^B*\Psi_\varphi)$$
$$+\frac i2\overline L*\gamma_a(\overrightarrow\partial^a
+\frac i2c_Aq^AA^{a(A)}\Sigma^A)*L
-\frac i2\overline L*\gamma_a(\overleftarrow\partial^a-
\frac i2c_Aq^AA^{a(A)}\Sigma^A)*L$$
$$+\frac i2\overline R\gamma_a*(\overrightarrow\partial^a R+
iq^0A^{a0}R)-\frac i2\overline
R\gamma_a*(\overleftarrow\partial^a R-iq^0A^{a0}R)$$
\begin{equation}\label{p4_2}
-\tilde h\overline L*\Psi_\varphi*R-\tilde h\overline
R*\bux\Psi\varphi*L)+m^2||\Psi_\varphi||^2-\frac f4||\Psi_\varphi||^4
\end{equation}

Здесь $\Sigma^A\in\O,A=0,1,\dots,7$ с модулем $\Psi_\varphi,\Psi\in\K$.

Лагранжиан свободных полей $\L_f$ имеет вид ($I,J,K,L=1,\dots,7$)
\begin{equation}\label{p4_3}
\L_f=-\frac14F^0_{ab}F^{ab(0)}-\frac1{16}\tr(\hat F_{ab}*\hat  F^{ab})=-\frac14F^0_{ab}F^{ab(0)}-\frac1{4}F_{ab}^KF^{ab(K)}$$ $$-\frac14f^{IJKL}q^{IJ}q^{KL}(A_a^IA_b^J-A_b^I
A_a^J)(A^{a(K)}A^{b(L)}-A^{b(K)}A^{a(L)})
\end{equation}
\begin{equation}\label{p4_4}
f^{IJKL}=\frac14\tr(\Sigma^I*\Sigma^J*\Sigma^K*\Sigma^l)
\end{equation}
$$\hat F_{ab}=\partial_b A_a^K\Sigma^K-\partial_a  A_b^K\Sigma^K+q^{IJ}(A_a^IA_b^J-A_b^IA_a^J)\Sigma^I*\Sigma^J$$
$$F^0_{ab}=\partial_b A^0_a-\partial_a A^0_b,\qquad
F_{ab}^k=\partial_\nu A_a^k-\partial_a A_b^k+\varepsilon^{IJK}q^{IJ}(A_a^IA_b^J-A_b^IA_a^J)$$

Величина $f^{IJKL}$ отражает неассоциативный характер  лагранжиана
свободных октонионных полей.

Введены левые и правые компоненты спиноров
\begin{equation}\label{p4_5}
\frac12(1+\gamma^5)\Psi=L,\qquad\frac12(1-\gamma^5)\Psi=R,\qquad\gamma^5=i\gamma^0\gamma^1\gamma^2\gamma^3
\end{equation}

$\Psi,\Psi_\varphi$ -- векторы из расширенного пространства
состояний $\K$, то есть это матрицы размерности $4\times4$. Элементы
этих матриц -- биспиноры. Числа $c_L,c_{L_0}$ обусловленны
нормировкой; $q^A$ -- заряды, с которым взаимодействуют поля $A_a^A$
с заряженными векторами состояния $\Psi$ и $\Psi_\varphi$; $q^{JK}$
-- заряды, определяющие вклад коммутаторов полей
$A_a^K,J,K=1,2,\dots,7$; $\gamma^a$ -- матрицы Дирака
($a=0,1,2,3;i=1,2,3$):
\begin{equation}\label{p4_6}
\gamma^0=\left(\matrix{I&0\cr0&-I}\right),\qquad
\gamma^i=\left(\matrix{0&\sigma^i\cr-\sigma^i&0}\right),\qquad \gamma^5=\left(\matrix{0&I\cr I&0}\right)
\end{equation}

Матрицы Дирака в (\ref{p4_2}) умножаются на векторы  из расширенного
пространства $\Psi\in\K$ тензорно.

По разновысотным греческим индексам осуществляется  суммирование с
метрическим тензором пространства Минковского $\eta_{ab}$ с
сигнатурой $(1,-1,$ $-1,-1)$, а по индексам одинаковой высоты --
просто суммирование.

В дальнейшем лагранжиан (\ref{p4_2}) предлагается называть
лагранжианом О-теории.

В \cite{Dor1} показано, что, если $R=e_R(x),\overline L=L^+\gamma^0$,
\begin{equation}\label{p4_7}
L=\frac{c_0}{\sqrt2}\left(\matrix{0\qquad(2i\sigma^1-2\sigma^2)\nu(x)
+(y_0I+\frac{2i}3\sigma^1+\frac23\sigma^2+i\sigma^3)e(x)\cr(-\frac
i8\sigma^1+\frac18\sigma^2)\nu(x)+(y_0I-\frac{3i}8\sigma^1-\frac38\sigma^2+
\frac{9i}{16}\sigma^3)e(x)\qquad0}\right)_L
\end{equation}
и, выбирая основное вакуумное состояние $\Psi_\varphi$ в виде
\begin{equation}\label{p4_8}
\Psi_\varphi=\Psi_0=\frac{m}{\sqrt{2f}}
\left(\matrix{0&i\sigma^3\cr0&I}\right)
\end{equation}
общее выражение лагранжиана О-теории для лептонного сектора имеет вид:
$$\L_{\o}=\L_f+
\frac{q^{(K)2}m^2}{2f}A_a^KA^{a(K)}+\frac{o^{IJ}m^2}{2f}A_a^{i}A^{a(j)}+\frac{g^{(1)2}m^2}{2f}B_a B^a-\frac{gg^{(1)}m^2}fA_a^3B^a$$
$$+\frac{g^{(1)}}2\overline\nu_L\gamma_a B^a\nu_L+
\frac{g^{(1)}}2\overline e_L\gamma_aB^a e_L+\frac g2\overline
e_L\gamma_aA^{a3}e_L-\frac g2\overline\nu_L\gamma_aA^{a3}\nu_L$$
$$-\frac g2\overline\nu_L\gamma_ae_L(A^{a1}-iA^{a2})-
\frac g2\overline e_L\gamma_a\nu_L(A^{a1}+iA^{a2})$$
$$+\frac i2(\overline e_L\gamma_a\partial^a e_L-\partial^a\overline
e_L\gamma_a e_L)+\frac i2(\overline\nu_L\gamma_a\partial^a\nu_L-
\partial^a\overline\nu_L\gamma_a\nu_L)+\frac{m^4}f$$
$$+\frac i2(\overline e_R\gamma_a\partial^a
e_R-\partial^a\overline e_R\gamma_a e_R)+g^{(1)}\overline
e_R\gamma_a B^a e_R-\frac{\sqrt2hm}{\sqrt f}(\overline
e_Le_R+\overline e_Re_L)$$
$$-q^4A^{a(4)}(\kappa_1\overline\nu_L\gamma_a\nu_L- \kappa_2\overline
e_L\gamma_ae_L)-\frac32q^6A^{a(6)}\overline e_L\gamma_ae_L$$
\begin{equation}\label{p4_9}
-\frac54(q^6A^{a(6)}+iq^5A^{a(5)})\overline\nu_L\gamma_a
e_L-\frac54(q^6A^{a(6)}-iq^5A^{a(5)})\overline e_L\gamma_a\nu_L
\end{equation}

Здесь введено обозначение $iA^{IJ}q^Iq^J=o^{IJ}$. Символ $A^{IJ}$  в
зависимости от модели снятия неассоциативности равен $\pm1,0$ и
антисимметричен по своим индексам. Числа $\kappa_1$ и $\kappa_2$
близки к числу десять \cite{Dor1}.

В этом лагранжиане имеется два члена, обусловленные
неассоциативностью  исходного лагранжиана. Во-первых, это выражение
\begin{equation}\label{p4_10}
\frac{o^{IJ}m^2}{2f}A_a^{I}A^{a(J)}
\end{equation}

Это выражение может принимать мнимые значения при  некотором выборе
порядка умножения октонионов, поэтому воспользуемся вероятностной
схемой снятия неассоциативности, которая обеспечит его нулевое
значение. По этой причине в дальнейшем исключим этот член из
рассмотрения.

Во-вторых имеется неассоциативный член (\ref{p4_4})  в лагранжиане
свободных полей. Снятие неассоциативности для этого члена приведёт к
появлению важных физических свойств лагранжиана О-теории, поэтому
обсуждение методов снятия его неаоссоциативности отложим до
соответствующего параграфа.

Нетрудно видеть, что, если положить старшие поля $A_a^K,K=4,5,6,7$ нулю, то мы приходим к лагранжиану Вайнберга-Салама -- Стандартной Теории слабых взаимодействий (СТ) в соответствующей калибровке. Собственно говоря, постановка задачи и состояла в выборе такого обобщения на $\K$-модуль, чтобы в частном случае младших полей получался бы лагранжиан именно СТ.

\section{Исследование лагранжиана октонионов}
1. Токовые части полей $A_\mu^k,k=0,1,2,3$ как и в  СТ указывают на
наличие векторных бозонов $Z^0,W$ и $W^*$, которые оказываются
массивными после исследовния их квадратичной части лагранжиана и
вакуумной, индуцированной $\Psi_0$. Этот факт является ожидаемым,
так как общий принцип получения лагрнжиана О-теории состоял в том,
чтобы при исключении из лагранжиана $\L_\o$ членов со старшими
полями $A_\mu^k,k=4,5,6,7$ мы приходили бы к лагранжиану СТ
\cite{Okun}.

2. По аналогии со СТ, во-первых, заметим, что токовая часть
\begin{equation}\label{p5_1}
q^4A^{\mu(4)}(\kappa_2\overline e_L\gamma_\mu e_L-\kappa_1\overline\nu_L\gamma_\mu\nu_L)
\end{equation}
лагранжиана (\ref{p4_9}) $\L_\o$ даёт основания ввести  нейтральное
векторное поле $C_\mu=A_\mu^4$, а вид квадратичных членов поля
$A_\mu^4$ позволяет ввести лагранжиан массивного векторного поля
$C_\mu$ ($m^2_C=q^{(4)2}m^2/f$)
\begin{equation}\label{p5_2}
\L_C=-\frac14(\partial_\mu C_\nu-\partial_\nu C_\mu+ \varepsilon^{4IJ}q^{IJ}(A_\mu^IA_\nu^J-A_\nu^IA_\mu^J))$$
$$(\partial^\nu C^\mu-\partial^\mu C^\nu+ \varepsilon^{4I'J'}q^{I'J'}(A^{(I')\mu}A^{(J')\nu}- A^{(I')\nu}A^{(J')\mu}))
+\frac{m^2_C}2C_\mu C^\mu
\end{equation}

3. Поле $A_\mu^7$ оказалось особенным -- оно не  содержит токов.
Учёт квадратичных членов по полю $A_\mu^7$ даёт основания ввести
массивное векторное поле $E_\mu=A_\mu^7$ ($m^2_E=q^{(7)2}m^2/f$) с
лагранжианом
\begin{equation}\label{p5_3}
\L_E=-\frac14(\partial_\mu E_\nu-\partial_\nu E_\mu+ \varepsilon^{7IJ}q^{IJ}(A_\mu^IA_\nu^J-A_\nu^IA_\mu^J))$$
$$(\partial^\nu E^\mu-\partial^\mu E^\nu+ \varepsilon^{7I'J'}q^{I'J'}(A^{(I')\mu}A^{(J')\nu}- A^{(I')\nu}A^{(J')\mu}))
+\frac{m^2_E}2E_\mu C^\mu
\end{equation}

4. Токовая часть
\begin{equation}\label{p5_4}
\frac54(q^6A^{\mu(6)}+iq^5A^{\mu(5)})\overline\nu_L\gamma_\mu
e_L+\frac54(q^6A^{\mu(6)}-iq^5A^{\mu(5)})\overline e_L\gamma_\mu\nu_L
\end{equation}
даёт основания ввести заряженный векторный бозон
\begin{equation}\label{p5_5}
D_\mu=\frac1{2q_D}(q^6A^{\mu(6)}-iq^5A^{\mu(5)})
\end{equation}
а учёт квадратичных членов $\L_\o$ позволяет ввести  массивное
векторное поле $D_\mu$ с лагранжианом (где $m_D^2=2m^2q_D^2/f$)
\begin{equation}\label{p5_6}
\L_D=-\frac14(\partial_\mu D_\nu-\partial_\nu D_\mu+ \varepsilon^{5IJ}q^{IJ}(A_\mu^IA_\nu^J-A_\nu^IA_\mu^J)+
\varepsilon^{6IJ}q^{IJ}(A_\mu^IA_\nu^J-A_\nu^IA_\mu^J))$$
$$(\partial^\nu D^\mu-\partial^\mu D^\nu+ \varepsilon^{5I'J'}q^{I'J'}(A^{(I')\mu}A^{(J')\nu}- A^{(I')\nu}A^{(J')\mu})+$$
$$\varepsilon^{6I'J'}q^{I'J'}(A^{(I')\mu}A^{(J')\nu}- A^{(I')\nu}A^{(J')\mu}))
+\frac{m^2_C}2C_\mu C^\mu
\end{equation}

Однако для этого необходимо считать, что $q^{(5)2}=q^{(6)2}=q_D^2=q_{D^*}^2$, поэтому $m_5=m_6=m_D=m_{D^*}$.

Но не всё так гладко! Имеется ещё одно интересное слагаемое
лагранжиана $\L_\o$
\begin{equation}\label{p5_7}
-\frac32q^6A^{\mu(6)}\overline e_L\gamma_\mu e_L
\end{equation}

В каком-то смысле это слагаемое также отвечает вектору тока. Действительно
\begin{equation}\label{p5_8}
-\frac32q^6A^{\mu(6)}\overline e_L\gamma_\mu e_L=
-\frac34(q_{D^*}D_\mu^*+q_DD_\mu)\overline e_L\gamma^\mu e_L
\end{equation}
но такой вид тока нарушает инвариантность лагранжиана относительно
глобального зарядового преобразования для векторных полей: если
\begin{equation}\label{p5_9}
D_\mu\to e^{iQ}D_\mu,\quad e_L\to e^{iQ}e_L\quad,\hbox{э. с.}
\end{equation}
то
\begin{equation}\label{p5_10}
(q_{D^*}D_\mu^*+q_DD_\mu)\overline e_Le_L\to (e^{-iQ}q_{D^*}D_\mu^*+e^{iQ}q_DD_\mu)e^{-iQ}\overline e_Le^{iQ}e_L$$
$$\ne(q_{D^*}D_\mu^*+q_DD_\mu)\overline e_Le_L
\end{equation}
Понятно, что потеря глобальной зарядовой инвариантности обусловлена
появлением массивного заряженного векторного бозона относительно
ненулевого вакуумного значения $\Psi_0$. Однако необычный вид
токового члена (\ref{p5_8}) требует его дополнительного
исследования. Пока можно утверждать: структура лагранжиана $\L_\o$
относительно вакуума $\Psi_0$ оказывается неинвариантной при
глобальном преобразовании заряда.

Некоммутативные члены лагранжианов $\L_C,\L_D,\L_E$ отличаются от
глюонных полей $SU(3)$-модели и в дальнейшем будем считать, что они
не оказывают влияния на исследуемую модель на расстояниях, больших
планковских.

\section{Решения для векторных бозонов в О-теории}
Запишем уравнения Эйлера-Лагранжа для векторных полей
$A_\mu^k,k=5,6$  лагранжиана $\L_\o$. Определяя напряжённости полей
$$F_{\mu\nu}^5=A_{\nu,\mu}^5-A_{\mu,\nu}^5,\quad
F_{\mu\nu}^6=A_{\nu,\mu}^6-A_{\mu,\nu}^6$$

Получим
\begin{equation}\label{p6_1}
F^{\mu(5)}_{\nu,\mu}+m_5^2A_\nu^5-\frac14f^{IJK5}q^{K5}q^{IJ}A^{\mu(I)}A_\nu^KA_\mu^J=0
\end{equation}
\begin{equation}\label{p6_2}
F^{\mu(6)}_{\nu,\mu}+m_6^2A_\nu^6-\frac14f^{I6KJ}q^{J6}q^{IK}A^{\mu(I)}A_\nu^JA_\mu^K=0
\end{equation}
где $m_i=m_D,i=5,6$. (В уравнениях (\ref{p6_1} -- \ref{p6_2}) не
выписана неабелева часть лагранжиана полей $\L_f$.)

Часть лагранжиана, отражающая его неассоциативный характер, осталась
в  $f^{IJKL}$. Ненулевые значения $f^{IJKL}$ в $\L_\o$ равны $\pm1$
только для некоторых комбинаций полей (\ref{p2_2}). Ограничиваясь в
(\ref{p6_1} -- \ref{p6_2}) ненулевыми значениями только для
$f^{4675}=-1$, которые выбираем из принципа минимакса модели,
получим (не будем учитывать перестановки индексов. Если же их
учитывать, то необходимо было бы ввести дополнительный множитель,
отвечающей за выбор снятия неассоциативности для каждой
перестановки):
\begin{equation}\label{p6_3}
F^{\mu(5)}_{\nu,\mu}+m^2_DA_\nu^5-q^{47}q^{56}A_\mu^4A_\nu^6A^{7(\mu)}=0
\end{equation}
\begin{equation}\label{p6_4}
F^{\mu(6)}_{\nu,\mu}+m^2_DA_\nu^6-q^{47}q^{56}A_\mu^4A_\nu^5A^{7(\mu)}=0
\end{equation}

Свёртка $A_\mu^4A^{\mu(7)}$ является скаляром и её можно считать
постоянной  величиной в некоторой малой области $\Omega_i$. В
квантовом представлении на такую свёртку можно смотреть как на
виртуальную пару массивных $C$ и $E$-бозонов. Так как в современном
эксперименте эти частицы ненаблюдаемы, то необходимо считать, что
они имеют очень большую массу. Будем также считать, что все
величины, входящие в (\ref{p6_3} -- \ref{p6_4}), подобраны так, что
$$m_D^2=q^{47}q^{56}A^{\mu(4)}A_\mu^7$$
откуда, в частности,
\begin{equation}\label{p6_5}
A^{\mu(4)}A_\mu^7=\frac{m_D^2}{q^{47}q^{56}}
\end{equation}
что можно допустить, учитывая массивность $C$ и $E$-бозонов.

Исследование частиц планковской массы пока наталкивается \cite{Dor3}
на  серьёзные экспериментальные трудности, поэтому ограничимся
только предположением (\ref{p6_5}). Тогда (\ref{p6_3}-\ref{p6_4})
представим как
\begin{equation}\label{p6_51}
\hat L^\mu_\nu A^{(5,6)}_\mu=m^2_D(A_\nu^{(5,6)}-A_\nu^{(6,5)})=0
\end{equation}
где введён оператор
\begin{equation}\label{p6_52}
\hat L^\mu_\nu=-\delta^\mu_\nu\Box+\partial_\nu\partial^\mu
\end{equation}

Определим новые функции:
\begin{equation}\label{p6_53}
\Theta^1_\nu=A^5_\nu+A^6_\nu,\qquad \Theta^2_\nu=A^5_\nu-A^6_\nu
\end{equation}

Как следует из (\ref{p6_51}) функции $\Theta^1_\nu,\Theta^2_\nu$
удовлетворяют следующим дифференциальным уравнениям:
\begin{equation}\label{p6_54}
\hat L^\mu_\nu\Theta^1_\nu=0,\qquad\hat L^\mu_\nu\Theta^2_\nu =2m^2_D\Theta^2_\nu
\end{equation}

Фактически речь идёт о решениях двух тапов: безмассовом $\Theta^1$
и массивном $\Theta^2$. Как будет видно из дальнейшего исследования
масса заряженного $D$ бозона имеет планковский порядок, поэтому
часть $\Theta^2$ очень высокочастотная и должна быстро убывать на
бесконечности, если речь идёт о стационарном решении, чего не
скажешь о безмассовом решении $\Theta^1$. Вспоминая определение
волновых функций заряженных $D_\nu$ и $D_\nu^*$ бозонов в
(\ref{p5_5})
\begin{equation}\label{p6_55}
D_\nu=\frac14(\Theta^1_\nu-\Theta^2_\nu)+\frac i4
(\Theta^1_\nu+\Theta^2_\nu)
\end{equation}
приходим к выводу, что на расстояниях больше планковских должно
быть
\begin{equation}\label{p6_56}
D_\nu\cong\frac{1+i}4\Theta^1_\nu
\end{equation}

Так как безмассовое решение соответствует условию
\begin{equation}\label{p6_6}
A_\nu^5=A_\nu^6
\end{equation}
то получим
\begin{equation}\label{p6_61}
D_\nu\cong\frac{1+i}2A^5_\nu
\end{equation}
Из (\ref{p6_3} -- \ref{p6_4}) следует, что тензор напряжённости
заряженных  безмассовых $D$-бозонов имеет вид:
\begin{equation}\label{p6_7}
F_{\nu\mu}^D=\partial_\mu D_\nu-\partial_\nu D_\mu,\quad F^{\mu\nu(D)}_{,\mu}=0
\end{equation}
который аналогичен тензору напряжёности электромагнитного поля. Так
как  физически подобных полей на опыте не обнаружено, то остаётся
положить
\begin{equation}\label{p6_71}
F_{\nu\mu}^D=\partial_\mu D_\nu-\partial_\nu D_\mu\equiv0
\end{equation}
Откуда получаем
\begin{equation}\label{p6_72}
D_\nu=\partial_\nu\alpha(x)
\end{equation}
где $\alpha(x)$ -- скалярная функция. Указанное решение, в
дополнении  с (\ref{p6_6}) и предположении о планковской массе
$D$-бозона действительно зануляет неассоциативные члены в
(\ref{p6_1}-\ref{p6_2})

В качестве решения (\ref{p6_71}) в сферической системе координат
можно взять  стационарное решение
\begin{equation}\label{p6_8}
A^5_\mu=A^6_\mu=(0,f(r,C),0,0)
\end{equation}
или однородное изотропное решение
\begin{equation}\label{p6_9}
A^5_\mu=A^6_\mu=(g(t,C),0,0,0)
\end{equation}
где $f(r,C)$ и $g(t,C)$ -- произвольные функци радиус-вектора $r$ и
времени $t$ соответственно с произвольной постоянной $C$.

С другой стороны, предполагая что в развиваемом в этой работе
приближении  О-теории, достаточно ограничиться неассоциативностью
только в слагаемом $f^{4567}$, следует учесть возможное влияние
этого члена в уравнении Эйлера-Лагранжа для бозонных $C$ и $E$
полей:
\begin{equation}\label{p6_10}
F^{\mu\nu(C)}_{,\mu}+m_C^2C^\nu=q^{47}q^{56}A^5_\mu A^{\mu(6)}E^\nu$$
$$F^{\mu\nu(E)}_{,\mu}+m_E^2E^\nu=q^{47}q^{56}A^5_\mu A^{\mu(6)}C^\nu
\end{equation}
(Здесь опять пренебрегаем всеми нелинейными членами по полям
$A_\mu^k$, кроме выписанных.)

Как было указано выше $A^5=A^6$ с быстрым убыванием к нулю, поэтому
мы приходим к уравнениям свободных массивных частиц.

До сих пор поля рассматривались исключительно как классические. При
вычислении свёртки $C_\mu E^\mu$ оказывается этого недостаточно.
Будем считать, что свёртка $C_\mu E^\mu$ находится как среднее
\begin{equation}\label{p6_11}
C_\mu E^\mu=<|\hat C_\mu\hat E^\mu|>\ne0
\end{equation}
где
\begin{equation}\label{p6_12}
\hat C_\mu=l_\mu^s(\hat c_s^+e^{-ikx}+\hat c_s e^{ikx}),\
\hat E^\mu=r_\mu^s(\hat e_s^+e^{-ikx}+\hat e_se^{ikx}),\ k^2=m^2
\end{equation}
и введены операторы рождения и уничтожения бозе-частиц и
предполагается  одинаковая масса частиц $C$ и $E$. Так как частицы
$C$ и $E$ различны и операторы $\hat c$ и $\hat e$ коммутируют, то
правая часть (\ref{p6_11}) по вакуумному состоянию равна нулю, но,
учитывая, что эти частицы, по предположению, образуют связанное
состояние, можно считать, что это не ноль. Например, эта свёртка
индуцирует ненулевую амплитуду перехода состояния из вакуума в
состояние $<1,1|=<0|\hat c\hat e$ и наоборот:
\begin{equation}\label{p6_13}
<0|\hat c\hat e(\hat c+\hat c^+)(\hat e^++\hat e)|0>=
<0|\hat c\hat c^+\hat e\hat e^+|0>\ne0$$
$$<0|(\hat c+\hat c^+)(\hat e^++\hat e)\hat c^+\hat e^+|0>=
<0|\hat c\hat c^+\hat e\hat e^+|0>\ne0
\end{equation}

Ненулевые значения на глюонном конденсате рассматривались ещё в
работе \cite{Shifman2}, где имеются ненулевое значение члена
$<\alpha_sG_{\mu\nu}^aG_{\mu\nu}^a>$, подтверждённое
экспериментально как нетривиальный вакуум глюонного кондесата. В
данной работе этот результат также постулироуется. Представляется,
что попытка численной оценки этой величины не разумна, так как
неизвестен вид нетривиальной структуры планковского вакуума.
Предполагается, что более разумным является проверка следствий
исследуемого подхода.

\section{Метод геометризации в О-теории}
Как было указано ранее, лагранжиан $\L_\o$ содержит вектор тока,
неинвариантный  относительно преобразования заряда (\ref{p5_10}).
Это обусловлено тем, что в лагранжиан $\L_o$ токи
\begin{equation}\label{p7_1}
q_DD_a^*(y)\overline e_L(y)\gamma^ae_L(y) \hbox{ и }
q_{D^*}D_a(y)\overline e_L(y)\gamma^ae_L(y)
\end{equation}
входят в виде суммы, а не в виде произведения. (Здесь
$y^a,a=0,1,2,3$ --  переменные плоского пространства Минковского.)
Более того, подставив этот лагранжиан в действие, получим что в
общем случае эти слагаемые оказываются определёнными в разных точках
области $U$
$$S'=\frac i2\int_U(\overline e\gamma^a(\partial_a+\frac{3i}2q_DD_a)e
-(\partial^a-\frac{3i}2q_{D^*}D^*_a\overline e\gamma_ae))d^4y$$
\begin{equation}\label{p7_2}
=\frac i2\int_U(\overline e\gamma^a(\partial_a+\frac{3i}2q_DD_a)ed^4y
-\frac i2\int_U(\partial_a-\frac{3i}2q_{D^*}D^*_a)\overline e\gamma^aed^4y
\end{equation}
(Учёт именно левых или правых токов обсудим в конце работы.)

С другой стороны, в исходном лагранжиане изначально входят не
частицы  $D$ и $D^*$ в разных точках, а только потенциал
$A^6_\mu(y)$. Поэтому пара частиц $D$ и $D^*$ должны быть и в
действии определена в одной точке. Для решения этой задачи проведём
процедуру геометризации.

Ограничимся случаем <<достаточно хороших связных областей $\Omega$>>
пространства Минковского и <<достаточно хорошо определённых
лагранжианов $\L_o$>>, для которых будут выполнены все перечисленные
ниже условия.

Рассмотрим достаточно большую область пространства Минковского
$\overline U=T\times\R^3$, когда $T$ и $\R^3$ компакты. Разобьём
область $\overline U$ произвольным образом на $n$ малых различных
компактных областей $\overline U_i\subset\overline U$ с границей
$\partial\overline U_i$ и внутренней областью $U_i=\overline
U_i\backslash\partial\overline U_i$
\begin{equation}\label{p7_3}
\bigcup_i\overline U_i=\overline U,\quad U_i\bigcap_{i\ne j}U_j=\varnothing,\quad i,j=1,\dots,n.
\end{equation}

В областях $U_i$ произвольным образом возьмём точки $A_i$ и  введём
локальные координаты $\xi_i$:
\begin{equation}\label{p7_5}
y_i=y(A_i)+\xi_i
\end{equation}

Тогда интервал между любыми двумя точками произвольной окрестности $U_i$ имеет вид
\begin{equation}\label{p7_6}
ds^2=(dy^0)^2-(dy^1)^2-(dy^2)^2-(dy^3)^2
\end{equation}

Введём многообразие $\overline M$. Разобьём $\overline M$ на $n$
областей $\overline M_i$ для которых
\begin{equation}\label{p7_7}
\bigcup_i\overline M_i=\overline M,\quad M_i\bigcap_{i\ne j} M_j=\varnothing,\quad i,j=1,\dots,n.
\end{equation}

Предположим, что для областей $M_i,\Omega_i$ и $U_i$ существуют
гомеоморфизмы
\begin{equation}\label{p7_8}
f_i:M_i\to\Omega_i\subset\R^4,$$
$$g_i:\Omega_i\to U_i,\quad p_i=f_i^{-1}\circ g_i^{-1}(A_i)
\end{equation}

Обозначим $x=x(p)=(x_0,\dots,x_3),p\in M, x_\mu(p)$ -- координаты
точек области $M$,  индуцируемые отображениями $f_i$ и дополнительно
считаем, что гомеоморфизмы $g_i,f_i$ и выбранные системы координат в
$U_i$ и в $\Omega_i$ таковы, что
\begin{equation}\label{p7_9}
y_a-y_a(p_i)=H_a^\mu(p_i)(x_\mu-x_\mu(p_i))
\end{equation}
где $H^\mu_a$ -- диагональная матрица.

Тогда квадратичная форма интервала относительно точки $p_i$ на многообразии $M_i$ имеет вид:
\begin{equation}\label{p7_10}
ds^2=H_0^2(dx^0)^2-H_1^2(dx^1)^2-H_2^2(dx^2)^2-H_3^2(dx^3)^2
\end{equation}

Запишем действие лагранжиана О-теории в виде интеграла Римана
\begin{equation}\label{p7_11}
S_\o=\int_U
\L_od^4y=\lim_{n\to\infty}\sum_{i=1}^n\L_o(A_i)
\Delta U_i,\quad\Delta U_i=\Delta y_0\Delta y_1\Delta y_2\Delta y_3
\end{equation}
или
\begin{equation}\label{p7_12}
S_\o=\lim_{n\to\infty}\sum_{i=1}^n\L_o(p_i)\sqrt{-g(p_i)}\Delta^4x,\quad \sqrt{-g(p_i)}=H_0H_1H_2H_3
\end{equation}

Выпишем спинорную часть действия
$$\Delta S_i'=(\frac i2(\overline e\gamma_\mu(H_\mu^{-1}\partial^\mu+\frac{3i}4(q^6A^{\mu(6)}-iq^5A^{\mu(5)}))e$$ $$-(H_\mu^{-1}\partial^\mu-\frac{3i}4(q^6A^{\mu(6)}+iq^5A^{\mu(5)})\overline e\gamma_ae))$$
\begin{equation}\label{p7_13}
-\frac58(q^6A^{\mu(6)}+iq^5A^{\mu(5)})\overline\nu\gamma_\mu
e-\frac58(q^6A^{\mu(6)}-iq^5A^{\mu5)})\overline e\gamma_\mu\nu) \sqrt{-g(p_i)}\Delta^4x
\end{equation}

Используя определение векторных полей $D$ и $D^*$ в (\ref{p5_5}) запишем действие (\ref{p7_13}) в виде
$$\Delta S_i'=(\frac i2(\overline e\gamma^\mu(\partial_\mu+\frac{3i}2q_DD_\mu)e
-(\partial^\mu-\frac{3i}2q_{D^*}D^*_\mu\overline e)\gamma_\mu e))$$
\begin{equation}\label{p7_14}
-\frac54q_{D^*}D^*_\mu\overline\nu\gamma_\mu
e-\frac54q_DD_\mu\overline e\gamma_\mu\nu)\Delta\Omega_i,\quad \Delta\Omega_i=\sqrt{-g(p_i)}\Delta^4x
\end{equation}

В дальнейшем пренебрежём распадом $D$-бозонов на лептоны, поэтому
исключим  из дальнейшего рассмотрения вторую строку в (\ref{p7_14}).

В малой окрестности $U_i$ некоторой точки пространства Минковского
рассмотрим  производную от биспинора дираковской частицы как
ковариантную производную в криволинейном пространстве, записанную в
тетрадном формализме в метрике (\ref{p7_10}) для которой тетрады
$H^\mu_a(p_i)$ подобраны таким образом, что
\begin{equation}\label{p7_15}
\Phi_\mu^2(p_i)=\frac34q^5A_\mu^5(p_i)\end{equation}
где $\Phi_\mu^2$ соответствует (\ref{p1_20}).

В новых координатах действие (\ref{p7_14}) принимает вид
\begin{equation}\label{p7_16}
\Delta S_i'=(\frac i2(\overline e\gamma^\mu(\partial_\mu +\frac34q^5A^{\mu(5)}+\frac{3i}4q^6A^{\mu(6)})e$$ $$-(\partial^\mu+\frac34q^5A^{\mu(5)}- \frac{3i}4q^6A^{\mu(6)}\overline e\gamma_\mu e)) \Delta\Omega_i
\end{equation}
что равносильно
\begin{equation}\label{p7_17}
\Delta S_i'=(\frac i2(\overline e\gamma^\mu(\nabla_\mu+\frac{3i}4q_{D^*}A^6_\mu)e-(\nabla_\mu-\frac{3i}4q_DA^6_\mu)\overline e\gamma^\mu e))\Delta\Omega_i
\end{equation}
или
\begin{equation}\label{p7_18}
\Delta S_i'=(\frac i2(\overline e\gamma^\mu\nabla_\mu e-\nabla_\mu\overline e\gamma^\mu\nabla_\mu e+\frac{3i}4(q_{D^*}A^6_\mu e+q_DA^6_\mu)\overline e\gamma^\mu e))\Delta\Omega_i
\end{equation}

Так как $A_\mu^6$ взято в одной точке $p_i$, а знаки взаимодействия
с  электроном частиц $D$ и $D^*$ противоположны, то
\begin{equation}\label{p7_19}
\Delta S_i'=\frac i2(\overline e\gamma^\mu\nabla_\mu e-\nabla_\mu\overline e\gamma^\mu e)\Delta\Omega_i
\end{equation}

Считая что подинтегральное выражение продолжаемо на всё $M$, получим
\begin{equation}\label{p7_20}
\Delta S'=\int_\Omega\frac i2(\overline e\gamma^\mu\nabla_\mu e-\nabla_\mu\overline e\gamma^\mu e)\sqrt{-g}dx^4
\end{equation}

Но  выписанная схема соответствует представлению интеграла Римана
для обычной задачи ОТО, поэтому перечисленные выше предположения для
некоторых задач выполнимы. Например, такая схема вычисления действия
подходит для решения Шварцшильда и Фридмана (\ref{p8_1}).

Необходимо иметь в виду, что ОТО кроме того, что предсказывает
явление  гравитации как эффект криволинейности пространства-времени
предлагает связь между материей и гравитацией через уравнения
Эйнштейна. Оказывается данное уравнение имеется и в лагранжиане
О-теории. Действительно, в лагранжиане О-теории имеется необычный
член
\begin{equation}\label{p7_21}
\Delta\L_f=\frac14f^{IJKL}q^{IJ}q^{KL}(A_a^IA_b^J-A_b^I
A_a^J)(A^{a(K)}A^{b(L)}-A^{b(K)}A^{a(L)})$$
$$=\tilde f^{IJKL}A_a^IA_b^JA^{a(K)}A^{b(L)}
\end{equation}
знак которого неопределён до объявления метода снятия неассоциативности.

Малые области $\Omega_i$ различаются между собой следующим образом:
имеются  области $\Omega_i$, в которых предполагается виртуальная
пара $D+D^*$-бозонов и есть области, в которых таких пар нет. В тех
областях, где виртуальная пара $D+D^*$ есть будем предполагать и
виртуальную пару $C\cdot E$-бозонов (вкладом младших полей
$A_\mu^B,B=0,1,2,3$ будем пренебрегать в виду их малой массы по
сравнению с массой старших полей $A_\mu^B,B=4,5,6,7$).

Вне области предполагаемой виртуальной пары $D+D^*$-бозонов
гравитационного  вакуума не образуется, поэтому среднее значение
виртуальной пары $C+E$ по такой области равно нулю.

Вспомним, что действие гравитационного поля в ОТО имеет вид \cite{Landau}:
\begin{equation}\label{p7_22}
S_g=-\frac1\kappa\int_\Omega R\sqrt{-g}d\Omega=-\frac1\kappa\int_\Omega G\sqrt{-g}d\Omega- \frac1{2\kappa}\int_\Omega\partial_\lambda(\sqrt{-g}w^\lambda)d\Omega
\end{equation}
где $w^\lambda$ -- некоторый вектор и
\begin{equation}\label{p7_23}
L_g=-\frac1\kappa G=-\frac1\kappa g^{\mu\nu}(\Gamma^\lambda_{\sigma\nu} \Gamma^\sigma_{\mu\lambda}-\Gamma^\lambda_{\sigma\lambda}\Gamma^\sigma_{\mu\nu})
\end{equation}
-- лагранжиан гравитационного поля.

Так как при нахождении уравнений движения дивергентные  члены можно
опустить, то не будем различать $G$ и $R$ под знаком интеграла.

Сопоставим  геометрический член неассоциативной части лагранжиана
(обоснованность такого предположения проверяется в случаях метрики
Шварцшильда и Фридмана) и учтём (\ref{p6_5}):
\begin{equation}\label{p7_24}
-\frac1\kappa\int_\Omega G\sqrt{-g}d\Omega=f^{4567}q^{47}q^{56}\int_\Omega A_\mu^4A^{\mu(7)}A_\nu^5A^{\nu(6)}\sqrt{-g}d\Omega=$$
$$=f^{4567}m_D^2\int_\Omega A_\mu^5A^{\mu(6)}\sqrt{-g}d\Omega
\end{equation}

\section{Уравнение Дирака на псевдоримановом пространстве}
Пусть $M$ -- некоторое псевдориманово многообразие, в каждой точке
$p\in M$ которого можно ввести координаты
$x(p)=(x^0,x^1,x^2,x^3)=x$, метрику
\begin{equation}\label{p8_1}
ds^2=g_{\mu\nu}dx^\mu dx^\nu
\end{equation}
и связность:
\begin{equation}\label{p8_2}
\Gamma^\mu_{\nu\lambda}=\frac12g^{\mu\kappa}(g_{\mu\kappa,\nu}+
g_{\nu\kappa,\lambda}-g_{\lambda\nu,\kappa})
\end{equation}

Тогда тензор Римана определим по формуле:
\begin{equation}\label{p8_3}
R^\tau_{\mu\nu\lambda}=\Gamma^\tau_{\mu\lambda,\nu}-
\Gamma^\tau_{\mu\nu,\lambda}+\Gamma^\tau_{\sigma\nu}\Gamma^\sigma_{\mu\lambda}-
\Gamma^\tau_{\sigma\lambda}\Gamma^\sigma_{\mu\nu}.
\end{equation}

Квадратичная форма (\ref{p8_1}) в окрестности каждой точки может
быть приведена к диагональному виду:
\begin{equation}\label{p8_4}
ds^2=H_0^2dx^{(0)2}-
H_1^2dx^{(1)2}-H_2^2dx^{(2)2}-H_3^2dx^{(3)2}
\end{equation}

Ассоциируя диагональные координаты с координатами физического
про\-странства-времени, считаем что метрика приведена к виду метрики
пространства Минковского $M_4$ в нормальной форме:
\begin{equation}\label{p8_5}
ds^2=c^2dt^2-dx^2-dy^2-dz^2=\eta_{ab}dx^adx^b
\end{equation}
(В этом параграфе будем различать греческие и латинские индексы,
полагая, что греческие индексы относятся к псевдориманову пространству, а
латинские -- к $M_4$.)

Пусть в каждой точке псевдориманова пространства $M$ определено
расслоение -- касательное пространство Минковского $M_4$ с метрикой
(\ref{p8_5}) и пусть
\begin{equation}\label{p8_6}
H_0dx^{(0)}=cdt,H_1dx^{(1)}=dx,H_2dx^{(2)}=dy,H_3dx^{(3)}=dz
\end{equation}

С другой стороны, в каждой точке псевдориманова пространства $M$
можно ввести  тетрады $h^a_\mu$, связывающие метрику псевдориманова
пространство и пространства Минковского $M_4$:
\begin{equation}\label{p8_7}
h^b_\mu h^\mu_\nu=\delta^b_a,\qquad h^{\mu(a)}h^\nu_a=g^{\mu\nu} \end{equation}

Всякий вектор $A_\mu$ псевдориманова пространства может быть
представлен покомпонентно в пространстве $M_4$ с помощью тетрад
$h_\mu^a$ как $A^a=A^\mu h^a_\mu$. Тогда $A^a$ -- вектор
относительно лоренц-преобразований. Обратным  преобразованием
$A^ah_a^\mu=A^\mu$ получим вектор в римановом пространстве
\cite{Landau}.

Перенесём параллельно вектора $A_\mu$ из точки $x_\nu$ в точку $x_\nu+\delta x_\nu$:
\begin{equation}\label{p8_8}
\delta A^\mu=\delta(A^ah^\mu_a)=\delta A^ah^\mu_a+A^a\delta h^\mu_a=\delta A^ah^\mu_a+A^ah^\mu_{a,\nu}\delta x^\nu=\Gamma^\mu_{\nu\lambda}A^\nu\delta x^\lambda \end{equation}

Учитывая $h^b_\mu h^\mu_{a,\nu}+h^b_{\mu,\nu} h^\mu_a=0$, находим
\begin{equation}\label{p8_9}
\delta A^b=\gamma^b_{ac}A^a\delta x^c,\qquad\gamma^b_{ac}=
h^b_{\mu;\nu}h^\mu_ah^\nu_c
\end{equation}
где $\gamma^b_{ac}$ -- коэффициенты вращения Риччи.

В работе Фока-Иваненко \cite{FokIvanenko} показано, что свободное
уравнение  Дирака в пространстве Минковского
\begin{equation}\label{p8_10}
(i\gamma^a\partial_a-m)e(x)=0
\end{equation}
записанное в $M_4$ как в касательном расслоении некоторого
псевдориманового  многообразия имеет вид
\begin{equation}\label{p8_11}
(ih^\mu_a\gamma^a(\partial_\mu-i\Phi_\mu^1-\Gamma_\mu)-m)e(x)=0
\end{equation}
где
\begin{equation}\label{p8_12}
\Gamma_a=h_a^\mu\Gamma_\mu=-\frac12\gamma_{abc}\sigma^{bc},\qquad \sigma^{bc}=\frac14[\gamma^b,\gamma^c],\qquad a,b=0,1,2,3
\end{equation}
и $\Phi_\mu^1$ -- произвольная вещественная функция.

Матрицы Дирака $\gamma^a,a=0,\dots,3$ удовлетворяют следующему
правилу умножения  с матрицами $\sigma^{ab}$:
\begin{equation}\label{p8_13}
\gamma^a\sigma^{bc}=\frac14\gamma^a[\gamma^b,\gamma^c]=\frac12\eta^{ab}\gamma^c-\frac12\eta^{ac}\gamma^b-\frac i2\varepsilon^{dabc}\gamma^5\gamma_d
\end{equation}

Следовательно
\begin{equation}\label{p8_14}
-\gamma^a\Gamma_a=
\frac14h^\mu_ah^\nu_bh_{(c)\nu;\mu}(\frac12\eta^{ab}\gamma^c-\frac12\eta^{ac}\gamma^b-\frac i2\varepsilon^{dabc}\gamma^5\gamma_d)
\end{equation}

Пусть метрика псевдориманова пространства имеет вид (\ref{p8_4}), но
в силу  её диагональности, а поэтому и диагональности тетрад
$h^\mu_a$ (набор тетрад выбираем именно такими) при различных
значениях $a,b,c$ получим
\begin{equation}\label{p8_15}
\Gamma^\lambda_{\mu\nu}h_{c\lambda}h^\mu_bh^\nu_c=\frac12g^{\mu\kappa}(g_{\mu\kappa,\nu}+
g_{\nu\kappa,\lambda}-g_{\lambda\nu,\kappa})h_{c\lambda}h^\mu_bh^\nu_c=0
\end{equation}
поэтому
\begin{equation}\label{p8_16}
-\gamma^a\Gamma_a=\frac14h^\mu_ah^{\nu(a)}h_{(c)\nu;\mu}\gamma^c-\frac14
h^{\mu(a)}h^\nu_bh_{(a)\nu;\mu}\gamma^b$$
$$=\frac14h^\mu_{c;\mu}\gamma^c+\frac14h^{\mu(a)}h^\nu_{b;\mu}h_{(a)\nu}\gamma^b=\frac12h^\mu_{c;\mu}\gamma^c
\end{equation}
и уравнение Дирака принимает вид \cite{SokolovIvanenko}:
\begin{equation}\label{p8_17}
(i\gamma^ah^\mu_a(\partial_\mu-i\Phi_\mu^1+\Phi_\mu^2)-m)\psi=0
\end{equation}
где введено обозначение
\begin{equation}\label{p8_18}
\Phi_\mu^2=\frac12\partial_\mu\left(\ln\frac{\sqrt{-g}}{H_\mu}\right)
\end{equation}

Производные в (\ref{p8_17}) берутся по переменным криволинейного
пространства  $M$. В некоторой окрестности точки $x(p)$
диагонализуем метрику (\ref{p8_1}) и введём координаты пространства
Минковского по формулам (\ref{p8_6}). Запишем (\ref{p8_17}) в этих
координатах:
\begin{equation}\label{p8_19}
(i\gamma^a(\partial_a-i\Phi_a^1+\Phi_a^2)-m)\psi=0
\end{equation}

Здесь $H_a=\delta^\mu_aH_\mu$ и
\begin{equation}\label{p8_20}
\Phi_a^2=\frac12\partial_a\left(\ln\frac{\sqrt{-g}}{H_a}\right)=
\frac12\delta^\mu_aH_\mu^{-1}\partial_\mu\left(\ln\frac{\sqrt{-g}}{H_\mu}\right)
\end{equation}

В частности, выпишем уравнение Дирака в метрике Шварцшильда. Для
этого запишем в сферически симметричной системе координат
относительно центра Земли метрику пространства-времени
\begin{equation}\label{p8_21}
ds^2=(1-\frac{r_g}r)dt^2-\frac{dr^2}{1-\frac{r_g}r}-
r^2(\sin^2\theta d\varphi^2+d\theta^2)
\end{equation}
где $r_g=2kM/R$ -- шварцшильдовский радиус.

Исключаем внешнее электромагнитное поле, тогда уравнение Дирака (\ref{p8_17})
относительно метрики Шварцшильда (\ref{p8_21}) принимает вид \cite{Wheeler1}
\begin{equation}\label{p8_22}
(\gamma^0\frac1f\partial_t+\gamma^r f\partial_r
-\gamma^r\frac{\vec\Sigma\cdot\hat{\vec L}}r-im)e=-\gamma^r(\frac12 f_{,r}+\frac1r(f-1))e,
\end{equation}
где $\vec L=\vec r\times\vec p$ -- оператор углового момента, $f^2=1-\frac{r_g}r$,
$$\gamma^r=\gamma^1\sin\theta\cos\varphi+
\gamma^2\sin\theta\sin\varphi+\gamma^3\cos\theta,\quad
\vec\Sigma=\left(\matrix{\vec\sigma&0\cr0&\vec\sigma}\right).$$

Подставляя $f$ в (\ref{p8_22}) находим, что в первом приближении
\begin{equation}\label{p8_23}
\frac12 f_{,r}+\frac1r(f-1)=-\frac{r_g}{4r^2}
\end{equation}

Тем самым найдена правая часть (\ref{p8_22}), которая выражает
отличие уравнения  Дирака в римановом пространстве от от уравнения
Дирака в искривлённом пространстве-времени \cite{Dor7}.

\section{Уравнение Дирака в пространстве Римана-Картана}
Ковариантная производная останется вектором и в том случае, если
символ Кристоффеля  содержит несимметричную по нижним индексам
часть. В этом случае псевдориманово пространство обобщается до
пространство Римана-Картана, в котором кручение, равное
антисимметричной части объекта связности,
\begin{equation}\label{p9_1}
Q^\lambda_{\mu\nu}=\frac12(\Gamma^\lambda_{\mu\nu}-\Gamma^\lambda_{\nu\mu})
\end{equation}
независимо наравне с псевдоримановой метрикой $g_{\mu\nu}$.

Тензор кручения имеет 24 независимые компоненты и может быть
разложен в сумму трёх  неприводимых частей
\begin{equation}\label{p9_2}
Q^\lambda_{\mu\nu}=\tilde Q^\lambda_{\mu\nu}+\frac13(\delta^\lambda_\mu  Q_\nu-\delta^\lambda_\nu  Q_\mu)+ \varepsilon_{\sigma\mu\nu\alpha} g^{\sigma\lambda}\breve Q^\alpha
\end{equation}
где $\tilde Q^\lambda_{\mu\nu}$ -- бесследовая часть тензора
кручения, $Q_\mu$ --  след тензора кручения
\begin{equation}\label{p9_3}
Q_\mu=Q^\lambda_{\mu\lambda}
\end{equation}
и $\breve Q^\alpha$ -- псевдослед тензора кручения
\begin{equation}\label{p9_4}
\breve Q_\alpha=\frac1{3!}\varepsilon_{\alpha\mu\nu\sigma}Q^{\mu\nu\sigma}
\end{equation}
(поднимаются и опускаются индексы с помощью метрического тензора $g^{\mu\nu}$).

Симметричная по нижним индексам часть объекта связности
$\Gamma^\lambda_{\mu\nu}$  в общем случае не обязательно согласована
с метрикой. Тем не менее в дальнейшем ограничимся случаем, когда его
симметричная часть определяется метрическим тензором
\begin{equation}\label{p9_5}
\dot\Gamma^\lambda_{\mu\nu}=\frac12(\Gamma^\lambda_{\mu\nu}+\Gamma^\lambda_{\nu\mu})=\frac12g^{\lambda\sigma}(g_{\nu\sigma,\mu}+g_{\mu\sigma,\nu}-g_{\mu\nu,\sigma})
\end{equation}
поэтому
\begin{equation}\label{p9_6}
\Gamma^\lambda_{\mu\nu}=\dot\Gamma^\lambda_{\mu\nu}+Q^\lambda_{\mu\nu}
\end{equation}

Но тогда коэффициент вращения Риччи (\ref{p8_9}) обобщается до
\begin{equation}\label{p9_7}
\gamma^b_{ac}=(h^b_{\mu,\nu}+\Gamma^\lambda_{\mu\nu}h^b_\lambda)h^\mu_ah^\nu_c=
(h^b_{\mu,\nu}+\dot\Gamma^\lambda_{\mu\nu}h^b_\lambda +Q^\lambda_{\mu\nu}h^b_\lambda)h^\mu_ah^\nu_c=$$
$$=(h^b_{\mu\tilde;\nu}+Q^\lambda_{\mu\nu}h^b_\lambda)h^\mu_ah^\nu_c
\end{equation}

Перейдём, в некоторой карте, к диагональному виду $M_4$
псевдоримановой метрики и ограничимся случаем,  когда бесследовая и
векторная части тензора кручения равны нулю. Тогда при различных
значениях $a,b,c$ в отличии от (\ref{p1_15}) получим
\begin{equation}\label{p9_8}
\Gamma^\lambda_{\mu\nu}h_{c\lambda}h^\mu_bh^\nu_c= Q^\lambda_{\mu\nu}h_{c\lambda}h^\mu_bh^\nu_c\ne0
\end{equation}
и
\begin{equation}\label{p9_9}
-\gamma^a\Gamma_a=\frac12h^\mu_{c\tilde;\mu}\gamma^c+i\gamma^5\gamma_d(-\frac18\varepsilon^{dabc}h^\mu_ah^\nu_bh_{(c)\lambda} Q^\lambda_{\mu\nu})
\end{equation}
и уравнение Дирака принимает вид:
\begin{equation}\label{p9_10}
(\gamma^a(H^a)^{-1}(\partial_a-i\Phi_a^1+\Phi_a^2+i\gamma^5\Phi_a^3)-m)\psi=0
\end{equation}
где $\Phi^{(d)3}=-\frac18\varepsilon^{dabc}h^\mu_ah^\nu_bh_{(c)\lambda} Q^\lambda_{\mu\nu}$.

\section{Решение Фридмана}
Покажем, что в плоском простанстве Фридмана имеется
самосогласованное решение  лагранжиана октонионов.

Рассмотрим однородную и изотропную Вселенную
\begin{equation}\label{p10_1}
ds^2=dx^{(0)2}-a^2(t)(dx^{(1)2}+dx^{(2)2}+dx^{(3)2})=
a^2(\eta)(d\eta^2-dl^2),
\end{equation}
где введено конформное время $dt=a(\eta)d\eta$, при этом ($\alpha,\beta=1,2,3$)
\begin{equation}\label{p10_2}
g_{00}=a^2(\eta),\quad g_{\alpha\beta}=a^2(\eta)\eta_{\alpha\beta}
\end{equation}

Находим все ненулевые компоненты символа Кристоффеля:
\begin{equation}\label{p10_3}
\Gamma^0_{00}=\frac{a'}{a^3},\quad\Gamma^0_{\alpha\beta}=-\frac{a'}{a^3}g_{\alpha\beta},\quad\Gamma^\alpha_{0\beta}=\frac{a'}a\delta^\alpha_\beta
\end{equation}
и вычислим величину
\begin{equation}\label{p10_4}
G=g^{\mu\nu}(\Gamma^\lambda_{\mu\nu}\Gamma^\kappa_{\lambda\kappa}- \Gamma^\lambda_{\mu\kappa}\Gamma^\kappa_{\nu\lambda})
=\frac{6{a'}^2}{a^4},
\end{equation}
где штрих означает производную по конформному времени.

Запишем уравнение Дирака в метрике Фридмана (\ref{p10_1})
\begin{equation}\label{p10_5}
(\gamma^\mu(H^\mu)^{-1}(\partial_\mu-i\Phi_j+\frac12
\partial_\mu\left(\ln\frac{\sqrt{-g}}{H^\mu}\right))-m)\psi=0
\end{equation}

Из (\ref{p7_15}) и (\ref{p10_5}) получаем, что
\begin{equation}\label{p10_6}
q_DA_\mu^5=((2da/(a^2d\eta),\vec 0)
\end{equation}
что согласуется с (\ref{p6_9}). Учтём (\ref{p6_5}), тогда
неассоциативный  по полям член лагранжиана (\ref{p4_4}) в
соответствии со схемой (\ref{p7_14}) приобретает вид
\begin{equation}\label{p10_8}
f^{4567}q^{56}q^{47}A^{(5)\mu}A^{(6)}_\mu A^{(4)\nu}A^{(7)}_\nu=\frac{q^{56}q^{47}}{q_D^2}f^{4567}A^{(4)\mu}A^{(7)}_\mu
\left(\frac{2da}{a^2d\eta}\right)^2=$$
$$=-\frac{m_D^2}{q_D^1}\frac{4a'{}^2}{a^4}
=-\frac6{\kappa}\left(\frac{da}{a^2d\eta}\right)^2
=-\frac6{\kappa}\frac{a'^2}{a^4}=-\frac1{\kappa}G=\L_g
\end{equation}
откуда получим $\kappa=3q_D^2/(2m_D^2)$ (вычисленния производим по
правилу  потенциальной схемы снятия неассоциативности). Тем самым
находим лагранжиан гравитационного поля во фридмановской моделе.

Рассмотрим правую часть (\ref{p6_10}) уравнений $C$ и $E$ бозонов.
\begin{equation}\label{p10_9}
F^{\mu\nu(C)}_{,\mu}+m_C^2C^\nu=
q^{56}q^{47}\frac{4a'^2}{q_D^2a^4}E^\nu$$
$$F^{\mu\nu(E)}_{,\mu}+m_E^2E^\nu=
q^{56}q^{47}\frac{4a'^2}{q_D^2a^4}C^\nu
\end{equation}

Если считать массы $C$ и $E$-бозонов большими, а скорость разбегания
Хаббла  малой, то при определённом выборе постоянной
$4q^{56}q^{47}/q_D^2$ действительно можно считать правую часть
(\ref{p10_9}) пренебрежимо малой.

\section{Решение Шварцшильда}
Рассмотрим пространство в котором имеется массивное сферически
симметричное  тело. Пусть это тело является источником октонионного
поля. На значительном расстоянии от этого тела электро-слабым
взаимодействием можно пренебречь, поэтому это тело может быть
источником разве что старших октонионных полей. В силу симметрии
задачи можно считать, если на большом расстоянии имеются октонионные
поля, источником которых является массивное тело, то они создаются
вектор-потенциалом $A_\mu^k=A_\mu^k(r),k=4,5,6,7$.

Пусть в этом пространстве движется, например, электрон.
Пространство, в  котором этот электрон движется -- по определению --
пространство Минковского. В сферически симметричных координатах
метрика в этом пространстве имеет вид:
\begin{equation}\label{p11_1}
ds^2=dt^2-dr^2-r^2(\sin^2\theta d\varphi^2+d\theta^2)
\end{equation}

Ограничимся левыми спинорами. Запишем уравнение движения электрона в
октонионном  поле массивного источника, предполагая, что с полями
$A^{4,7}_\mu$ он не взаимодействует и исключая поле $A_\mu^6$
\begin{equation}\label{p11_2}
(i\gamma^0\partial_0+i\gamma^r(\partial_r-\vec\Sigma\cdot\hat{\vec L}-\frac{3i}4q^6A^6_r-\frac34q^5A^5_r)+m)\psi=0
\end{equation}
Здесь введено обозначение
$$\gamma^r=\gamma^1\sin\theta\cos\varphi+
\gamma^2\sin\theta\sin\varphi+\gamma^3\cos\theta,\quad
\vec\Sigma=\left(\matrix{\vec\sigma&0\cr0&\vec\sigma}\right)$$
где $\hat{\vec L}=\hat{\vec r}\times\vec p$ -- оператор углового момента \cite{Wheeler1}.

В соответствии с общей схемой геометризации О-теории получим:
\begin{equation}\label{p11_3}
\frac12\partial_r\left(\ln\frac{\sqrt{-g}}{H^r}\right))=-\frac34q^5A^5_r
\end{equation}

Введём общий вид стационарной сферически-симметричной метрики,
полученной в результате  геометризации О-теории
\begin{equation}\label{p11_4}
ds^2=H_0^2(r)dt^2-H^2_1(r)dr^2-r^2(\sin^2\theta d\varphi^2+d\theta^2)
\end{equation}
Нетрудно убедиться, что $g(r)=-H_0^2H_1^2$. Поэтому имеется одна
ненулевая компонента  вектора $A_\mu^5$.
\begin{equation}\label{p11_5}
-\frac34q^5A^5_r=\frac12\partial_r\left(\ln\frac{\sqrt{-\bar g}}{H^r}\right))=\frac{H_{0,r}}{2H_0},\qquad A^5=(0,A^5_r,0,0)
\end{equation}

Известно, что в случае слабого поля \cite{Landau} достаточно
исследовать только $H_0$.  Соответствия геометрической модели и
ньютоновского приближения обеспечивается выполнением равенства
\begin{equation}\label{p11_6}
H_0^2=g_{00}=1-r_g/r=f^2
\end{equation}
где $r_g$ -- гравитационный радиус. При этом $r>>r_g$, то есть
$r_g/r=\alpha(r)$ --  бесконечно малая по $r$. Будем искать $H_1$
как $1+C/r^n,n>1$.

\begin{equation}\label{p11_7}
H_1=1+C/r^n=1+\beta(r),\quad n>1
\end{equation}

Найдём лагранжиан гравитационного поля (\ref{p7_13}) для метрики (\ref{p11_4})
\begin{equation}\label{p11_8}
G_g=g^{\mu\nu}(\Gamma^\lambda_{\mu\nu}\Gamma^\kappa_{\lambda\kappa}- \Gamma^\lambda_{\mu\kappa}\Gamma^\kappa_{\nu\lambda})
=\frac2{r^2H_1^2}+\frac{4H_{0,r}}{rH_1^2H_0}-\frac{H_{0,r}H_{1,r}}{H_1^3H_0}
\end{equation}

Так как для плоской метрики (\ref{p11_1}) лагранжиан гравитационного
поля (\ref{p7_13}) равен $2/r^2$, то лагранжиан гравитационного поля
в О-теории, как это следует из (\ref{p7_14}) и (\ref{p11_5}), имеет
вид:
\begin{equation}\label{p11_9}
\frac1\kappa G_g=\frac1\kappa G_{pl}+\frac{2m_D^2}{3q_D^2}\frac32q_D^2A_\mu^5A^{(5)\mu} =\frac1\kappa(\frac2{r^2}+\frac{2H_{0,r}^2}{3H_0^2})
\end{equation}

Учитывая, что $\alpha(r),\beta(r)$ -- бесконечно малые, убеждаемся,
что  согласования (\ref{p11_8}) и (\ref{p11_9}) можно добиться
только при условии, если
\begin{equation}\label{p11_10}
H_0H_1=1,\qquad g_{11}=(1-r_g/r)^{-1}
\end{equation}

В результате приходим к метрике Шварцшильда
\begin{equation}\label{p11_11}
ds^2=(1-\frac{r_g}r)dt^2-\frac{dr^2}{1-\frac{r_g}r}-
r^2(\sin^2\theta d\varphi^2+d\theta^2)
\end{equation}

С другой стороны, учитывая левые и правые токи в лагранжиане
О-теории  уравнение Дирака можно записать в виде:
\begin{equation}\label{p11_12}
(\gamma^\mu(\partial_\mu+i\frac34q^6A_\mu^6+i\frac34q^6A_\mu^6\gamma^5 +\frac34q^5A_\mu^5+\frac34q^5A_\mu^5\gamma^5)+im)e(x)=0
\end{equation}

Это уравнение в длинной производной имеет два вектора и два
псевдовектора.  Пусть $\epsilon_{g_i}=0,1$, тогда длинная
производная в уравнение Дирака (\ref{p11_12}) может быть
представлена так
\begin{equation}\label{p11_13}
\partial_\mu+i\epsilon_{g_1}\frac34q^6A_\mu^6+i\epsilon_{g_2}\frac34q^6A_\mu^6\gamma^5+ \epsilon_{g_3}\frac34q^5A_\mu^5 +\epsilon_{g_4}\frac34q^5A_\mu^5\gamma^5
\end{equation}

Пусть $\epsilon_{g_3}=1,\epsilon_{g_i}=0,i=1,2,4$. Тогда мы имеем
аналогию  удлинения производной -- как тетрадного представления
уравнения Дирака, записанного в римановом пространстве в
диагональной метрике (\ref{p1_6})
\begin{equation}\label{p11_14}
(\gamma^\mu(\partial_\mu+\frac34q^5A_\mu^5)+im)e(x)=0
\end{equation}

Сравнивая (\ref{p1_23}) и (\ref{p11_3}) и учитывая (\ref{p5_6}) и
(\ref{p6_6}),  можно найти значение октонионных полей
\begin{equation}\label{p11_15}
q^5A_\mu^5=q^6A_\mu^6=(0,-\frac{r_g}{3r^2},0,0)
\end{equation}

В единицах СИ, получим:
\begin{equation}\label{p11_16}
q^5A^5_r=\frac{2kM\hbar}{3c^3r^2}=\frac{2g\hbar}{3c}= 9.8\cdot10^2cm/c^2\cdot10^{-27}erg\cdot c\cdot0.22\cdot10^{-10}c/cm$$
$$=0.2\cdot 10^{-34}erg\approx10^{-23}eV
\end{equation}

Пусть $\epsilon_{g_2}=1,\epsilon_{g_i}=0,i=1,3,4$. Тогда мы имеем
аналогию  удлинения производной -- как тетрадного представления
уравнения Дирака, записанного в пространстве Римана-Картана
\begin{equation}\label{p11_17}
(\gamma^\mu(\partial_\mu+i\gamma^5\frac34q^6A_\mu^6)+im)e(x)=0
\end{equation}
где $q^6A^6_\mu$ указано в (\ref{p11_15} -- \ref{p11_16}).

Заметим, ччто полученные выше выражения для вектор-потенциалов $A^5,A^6$ найдены в приближении слабого поля для проверки согласования предлагаемой модели О-теории и ОТО. По формуле (\ref{p11_3}) следует следующая точная формула:
\begin{equation}\label{p11_18}
A^6_r=-\frac{r_g}{3r^2(1-r_g/r)}
\end{equation}
которые даёт более существенные значения на расстояниях, сравнимых с
гравитационным радиусом.

\section{Нерелятивисткий электрон\\ в пространстве Римана-Картана}
Выделим спинорные компоненты в биспиноре, считая поле стационарным
\begin{equation}\label{p12_1}
e(x)=e^{-i\varepsilon t/h}\left(\matrix{\varphi(r)\cr\xi(r)}\right)
\end{equation}
тогда, обозначая $T^c=(0,\frac{r_g}{4r^2},0,0)$, получим (\ref{p11_17})
\begin{equation}\label{p12_2}
\left\{\matrix{\displaystyle\varepsilon\varphi+c\vec\sigma\vec p\xi-mc^2\varphi=-\hbar c\sigma^rT_r^c\varphi\cr\cr\displaystyle
\varepsilon\xi+c\vec\sigma\vec p\varphi+mc^2\xi=-\hbar c\sigma^rT_r^c\xi}\right.
\end{equation}
или, вводя $E'=\varepsilon-mc^2$, имеем
\begin{equation}\label{p12_3}
\left\{\matrix{\displaystyle(E'+\hbar c\vec\sigma\vec T^c)\varphi=-c\vec\sigma\vec p\xi\cr\cr\displaystyle(E'+2mc^2+\hbar c\vec\sigma\vec T^c)\xi=-c\vec\sigma\vec p\varphi}\right.
\end{equation}

Откуда для $|E'+T_r^c|<<2mc^2$
\begin{equation}\label{p12_4}
\xi=-(E'+2mc^2+\hbar c\vec\sigma\vec T^c)^{-1}c\vec\sigma\vec p\varphi\approx-\frac1{2mc}\vec\sigma\vec p\varphi
\end{equation}

Подставим (\ref{p12_4}) в первое уравнение системы (\ref{p12_3}),
тогда для  компоненты $\varphi$, биспинора $e(x)$ получим:
\begin{equation}\label{p12_5}
E'\varphi\approx(\frac1{2m}(\vec\sigma\vec p)^2-\hbar c\vec\sigma\vec T^c)\varphi
\end{equation}

Из квантового уравнения (\ref{p12_5}) следует, что нерелятивисткая
частица,  обладающая спином $\vec\sigma$, в поле Земли имеет энергию
\begin{equation}\label{p12_6}
E=\frac{\vec p^2}{2m}+V_g+V_c,
\end{equation}
где $V_g$ -- гравитационная энергия частицы и $V_c$ -- потенциальная
энергия  кручения Картана. При этом \cite{Dor5}
\begin{equation}\label{p12_7}
V_c=-\hbar c\vec\sigma\vec T^c=-\frac{\hbar r_gc}{4r^3}(\vec\sigma\vec r)
\end{equation}

Имеются эксперименты, в которых измеряется поправки к уравнению
Дирака  на поверхности Земли. Этот же эффект может быть вычислен по
нарушению преобразований Лоренца \cite{Russell}. На данный момент
расчёты показывают, что эффект, обусловленный кручением не наблюдаем
по техническим возможностям. Здесь показано, что этот эффект в
модели О-теории имеет значение $10^{-23}eV$, а наилучшие измерения
дают оценку $10^{-21}eV$. На самом деле оценка в $10^{-21}eV$
\cite{Heck} не соответствует эффекту, исследуемому в данной работе,
так как в \cite{Heck} исследовались неполяризованные электроны, в то
время как здесь имеются в виду левополяризованные электроны,
находящиеся в дублете с нейтрино.

Учитывая результат (\ref{p11_18}) получим следующее значение энергии Картана
\begin{equation}\label{p12_71}
V_c=-\hbar c\vec\sigma\vec T^c=-\frac{\hbar r_gc}{4r^3(1-\frac{r_g}r)}(\vec\sigma\vec r)
\end{equation}
которая оказывается очень большой для на поверхности <<чёрных дыр>>.
Из (\ref{p12_71})  следует важное экспериментальное следствие:
спиноры, излучаемые невращающейся чёрной дырой должны быть
поляризованы, при этом их спин должен быть направлен по радиусу от
звезды.

\section{Геометризация слабых взаимодействий в О-теории}
Будем рассматривать решения, соответствующие лептонному и кварковому
секторам как решения вне <<чёрной дыры>> и внутри. Вне <<чёрной
дыры>> имеем лептонное решение, а внутри -- кварковое. В результате,
внутри <<чёрной дыры>> приходим к дуплету u, d-кварков. Однако на
опыте имеются три поколения кварков. Возникает вопрос о том, почему
необходимо рассматривать именно три поколения и имеются ли ещё
покления? Оказывается действительно необходимо брать три поколения.
Дело в том, что теория слабых взаимодействий в нашем случае
получается как теория представления на алгебре октонионов. Алгебра
октонионов может быть реализована как групповая симметрия $G_2$.
Групповая симметрия $G_2$ имеет реализацию симметрии $SU(3)$
\cite{India}, которая имеет три подалгебры $SU(2)$. В дальнейшем
пока ограничимся одним поколением кварков.

Выберем состояние $\Psi$ для $u$ и $d$-кварков (решение находим по
аналогии с  лептонным сектором, но другим оператором гиперзаряда
$Y$)
\begin{equation}\label{p14_1}
\Psi=\sum_{i=0}^9\alpha_iu+\sum_{i=0}^9\beta_id,\quad
Y_{u,d}=\left(\matrix{-\frac13&0\cr0&\frac23}\right)
\end{equation}
где $\alpha_i$ и $\beta_i$ -- некоторые коэффициенты, для которых
кватернионный  сектор вектора состояний $\Psi$ соответствует
лагранжиану слабых взаимодействий.

Так же как и в случае слабых взаимодействий запишем полный
лагранжиан О-теории,  пренебрегая различием между левыми и правыми
частицами ($A,B=0,1,\dots,7$):
$$\L_\o=\L_f+(\partial_a\bux\Psi\varphi-
\frac i2q^AA_a^A\bux\Psi\varphi*\Sigma^A)
*(\partial^a\Psi_\varphi+
\frac i2q^BA^{a(B)} \Sigma^B*\Psi_\varphi)$$
$$+\frac i2\overline\Psi*\gamma_a(\overrightarrow\partial^a
+\frac i2c_Aq^AA^{a(A)}\Sigma^A)*\Psi
-\frac i2\overline L*\gamma_a(\overleftarrow\partial^a-
\frac i2c_Aq^AA^{a(A)}\Sigma^A)*\Psi$$
\begin{equation}\label{p14_2}
-\tilde h\overline \Psi*\Psi_\varphi*\Psi-\tilde h\overline
\Psi*\bux\Psi\varphi*\Psi)+m^2||\Psi_\varphi||^2-\frac f4||\Psi_\varphi||^4
\end{equation}

Выбирая основное состояние скалярного поля
$\Psi_\varphi=\Psi_{\varphi_0}$ из  (\ref{p4_8}), выпишем некоторые
члены лагранжиана $\L_o$:
$$\L_\o'=
\frac{q^{(5)2}m^2}{2f}A_a^5A^{a(5)}+\frac{q^{(6)2}m^2}{2f}A_a^6A^{a(6)}
+q^6A^{a(6)}(\kappa_3\overline u\gamma_a u+\kappa_4\overline d\gamma_ad)$$
\begin{equation}\label{p14_3}+\kappa_5(q^6A^{a(6)} +iq^5A^{a(5)})\overline u\gamma_ad+\kappa_5(q^6A^{a(6)}-iq^5A^{a(5)})\overline d\gamma_au$$
$$+f^{1256}q^{12}q^{56}A_a^1A^{a(2)}A_b^5A^{b(6)}
\end{equation}

Подставим групповую постоянную слабых взаимодействий $q^{12}=g$
Таким образом, как видно из последнего члена (\ref{p14_3}),
массивные заряженные  векторные бозоны $D$ и $D^*$ по аналогии с
моделью гравитации в представлении О-теории, индуцируют гравитацию,
которую можно назвать сильной гравитацией, если $A_a^1A^{a(2)}\ne0$,
но уже с другой постоянной
\begin{equation}\label{p14_4}
\kappa_s=|\frac{gq^{56}}{q_D^2}A_a^1A_a^2|
\end{equation}

Также как и среднее значение $<C_\mu E^\mu>\ne0$ здесь постулируется
ненулевое значение $<A_a^1A^{a(2)}>\ne0$. Однако в данном случае мы
имеем дело не с планковскими массами, а с теперь уже достаточно
рядовой массой заряженного векторного бозона, относительно
вакуумного значения в глюонном конденсате. Известные оценки для
$\alpha_sG_{\mu\nu}^aG_{\mu\nu}^a\approx0.02 Gev^4$ \cite{Shifman2},
дают требуемый порядок для (\ref{p14_4}). В данной моделе роль
нетривиального вакуума \cite{Dor4} может играть и гравитация. Тем
неменее аккуратное исследование этого вопроса, когда исходная
алгебра неассоциативна и поэтому можно говорить о квантовании лишь
на некоторой моделе предлагается здесь не обсуждать, а
постулировать.

Тогда решение для устойчивого ядра можно рассматривать как решение
типа  <<чёрной дыры>> в метрике Шварцшильда, объясняя невидимость
свободных кварков как решений внутри <<чёрной дыры>>, которая <<не
имеет волос>>, и поэтому оказывается невозможным извлечение
какой-либо информации, кроме информации о её массе, заряде и
собственном моменте количества движения.

Одим из первых выводов рассматриваемой модели является невозможность
участвовать в сильных взаимодействиях частиц массой сравнимой с
массой векторного бозона, так как на комптоновской длине волны
массивной частицы среднее значение $A_a^1A^{a(2)}$ нельзя считать
отличным от нуля. Такой вывод возможно подтверждается отсутствием
топпония, так как масса топ кварка больше массы заряженных векторных
$W$-бозонов.

Таким образом по аналогии с ОТО в представлении неассоциативной
алгебры  считаем, что последний член лагранжиана (\ref{p14_3})
записан в искривлённом пространстве-времени с коэффициентом связи
$\kappa_s$, то есть
\begin{equation}\label{p14_5}
f^{1256}q^{12}q^{56}A_a^1A^{a(2)}A_b^5A^{b(6)}\to-\frac1\kappa_sR
\end{equation}
где $R$ -- кривизна Римана.

Возникает вопрос о допустимости интерпретации одного и того же поля
$A_b^5A^{b(6)}$ в  смысле гравитации и в смысле сильных
взаимодействий, так как безмассовость поля $A^5$ была обусловлена
удачным подбором среднего значения  $C_\mu E^\mu$. Было показано,
что решения (\ref{p6_8} -- \ref{p6_9}) допустимы и обладают
геометрической природой, являюсь решением некоторой совместной
системы, то будем говорить о разных решениях $f,g$ из (\ref{p6_8} --
\ref{p6_9}) как о решениях с разными постоянными $C$.

\section{Модель элементарной частицы в О-теории}
Если элементарная частица не имеет спина и заряда, то в соответствии
с  развиваемым здесь подходом можно говорить, что статическое
состояние таких частиц индуцирует геометрию Римана с метрикой
\begin{equation}\label{p15_1}
ds^2=(1-\frac{r_s}r)dt^2-\frac{dr^2}{1-\frac{r_s}r}-r^2(\sin^2\theta d\varphi^2+d\theta^2)
\end{equation}
где $r_s$ -- гравитационный радиус сильных взаимодействий.

В соответствии с ОТО
\begin{equation}\label{p15_2}
r_s=2M=\frac{2\kappa_sm}{c^2}
\end{equation}
где $\kappa_s$ -- постоянная сильного взаимодействия, аналогичная
гравитационной  постоянной. Эта постоянная приводит к появлению силы
притяжения, аналогичной гравитационному притяжению с нерелятивистким
потенциалом (как и ньютоновский потенциал):
\begin{equation}\label{p15_3}
\varphi_s=-\kappa_s\frac mr
\end{equation}

Однако ввиду короткодействующего вида потенциала заряженных
массивных векторных  бозонов этот потенциал незаметен на больших
расстояниях (расстояниях, больших размера атомного ядра).

Найдём оценочное значение постоянной связи сильного взаимодействия
$k_s$ на основе  размера нейтрона $r_{s(n)}=2F=2\cdot10^{-13}$ см и
его массы $m_{n}=1.7\cdot10^{-24}$ г:
\begin{equation}\label{p15_4}
\kappa_s=\frac{c^2r_s}{2m_{n}}=5.4\cdot10^{31}\frac{cm^3}{gram\cdot sec^2}
\end{equation}
что в $10^{39}$ раз больше гравитационной постоянной.

В том случае, когда ядро обладает зарядом и вращательными степенями
свободы его гравитационный радиус может измениться. Действительно,
будем считать, что по аналогии с моделью заряженной вращающейся
<<чёрной дыры>> в ОТО, в теории сильных взаимодействий
рассматриваемой модели сильной гравитации метрика имеет вид
геометрии Керра-Ньюмана
\begin{equation}\label{p15_5}
ds^2=(\Delta/\rho^2)(dt-a\sin^2\theta d\varphi)^2-$$
$$(\sin^2\theta/\rho^2) ((r^2+a^2)d\varphi-adt)^2-(\rho^2/\Delta)dr^2-\rho^2d\theta^2
\end{equation}
$$\Delta=r^2-2Mr+a^2+\bar Q^2,\quad \rho^2=r^2+a^2\cos^2\theta$$
записанную в координатах Боейра-Линдквиста \cite{Wheeler2}.

Тогда радиус ядра определяется формулой
\begin{equation}\label{p15_6}
r^2-2Mr+a^2+\bar Q^2=0
\end{equation}
Здесь $a=L/(mc)$ -- момент <<чёрной дыры>>, приходящийся на единицу
её массы,  и $\bar Q$ -- электрический заряд в гравитационных
единицах. Нетрудно получить связь между обычным зарядом $Q$
выраженным в единицах $SGS_q$ (когда электрическая напряжённость
$E_q$ в центрально-симметричном поле определяется формулой
$E_q=Q/r^2$) и зарядом $\bar Q$ в гравитационных единицах
\cite{Tolman}:
\begin{equation}\label{p15_7}
\bar Q^2=\frac{4\pi\kappa_sQ^2}{c^4}
\end{equation}

Сделаем оценку величины заряда $\bar Q$ для атома водорода
$Q=4.8\cdot10^{-10}$ ед. СГСЭ (Учтем, что 1 ед. СГСЭ=1 $din\cdot
cm^2$):
\begin{equation}\label{p15_8}
\bar Q^2=\frac{4\pi\cdot5.4\cdot10^{31}cm^3\cdot(4.8\cdot10^{-10} SGS_Q)^2} {gram\cdot sec^2(3\cdot10^{10}cm/sec)^4}=1.96\cdot10^{-28}cm^2
\end{equation}

Оценим угловой момент $a$ элементарной частицы. В случае протона
получим
\begin{equation}\label{p15_9}
a=\frac {L_p}{m_pc}=\frac\hbar2(1.7\cdot10^{-24}gram\cdot3\cdot10cm/sec)^{-1}=10^{-14} cm
\end{equation}

Таким образом вклад углового момента и заряда на размер протона
оказался  примерно одинаков: $\bar Q_p/r_p=0.14,a_p/r_p=0.1$. При
этом следует иметь ввиду, что с ростом углового момента элементарной
частицы растёт и её масса по формуле
\begin{equation}\label{p15_10}
L=\alpha+\beta M^2
\end{equation}
поэтому в дальнейшем будем пренебрегать влиянием на размер
элементарной частицы и её зарядом и её угловым моментом.

Рассмотрим внутреннее решение Шварцшильда --
центрально-симмет\-ричное решение  под горизонтом при $r<r_s$. Один
из подходов исследования такого решения -- переход в систему отсчёта
падающих частиц на центр. Тогда, как известно, переходя к
координатам Леметра, мы убеждаемся, что никаких особенностей при
переходе черех границу $r=r_s$ не возникает, но мы никогда не сможем
вернуться назад. На самом деле решение Шварцшильда достаточно
удовлетворительное и для наших целей. Для этого при малых временах
рассмотрим преобразования вида \cite{Novikov}
\begin{equation}\label{p15_11}
ct\to r,\quad r\to ct,\quad r\in(0;+\infty),\quad t\in(0;r_g/c)
\end{equation}

В результате получим метрику внутри <<чёрной дыры>>
\begin{equation}\label{p15_12}
ds^2=\frac{c^2dt^2}{\frac{r_s}{ct}-1}-(\frac{r_s}{ct}-1)dr^2-c^2t^2(\sin^2\theta d\varphi^2+d\theta^2)
\end{equation}

Если $t\to0$, то член при $d\Omega$ пропадает. Следовательно  внутри
<<чёрной дыры>> геометрия оказывается двумерной. Исключая угловые
координаты в (\ref{p15_4}), получим в представлении собственного
времени $T=\frac23t\sqrt{t/t_s}$
\begin{equation}\label{p15_13}
ds^2=c^2dT^2-\left(\frac{2t_s}{3T}\right)^{2/3}dr^2
\end{equation}

При дальнейшем уменьшении времени член при $dr$ увеличивается,  что
можно понимать как расширение пространства-времени. В таком мире
кварки должны бесконечно удаляться друг от друга.

Заметим, что смысл временной переменной внутри <<чёрной дыры>>
совсем другой, нежели в обычном представлении, поэтому можно
рассматривать малые $t$ по сравнению с $t_s$, на которых динамика
под горизонтом Шварцшильда сжимающаяся, то есть $t$ уменьшается,
придавая кваркам релятивисткие скорости. В этом случае
схлопывающаяся геометрия приобретает смысл сил притяжения . Внутри
<<чёрной дыры>> \cite{Novikov}, в случае решения Шварцшильда,
пространство пустое и всякое массиное тело под горизонтом
Шварцшильда <<схлопывается>> за конечное время, поэтому свободного
кварка видимо существовать не может. В предлагаемом подходе будем
предполагать, что на малых расстояниях частица описывается квантовой
механикой и между парой частиц могут возникнуть квантовомеханические
стационарные состояния.

Отметим, что пространство внутри <<чёрной дыры>> трёхмерно.
Неизотропность по радиальной  и угловой переменным можно понимать
так, что материя, индуцирующая метрику, распределена либо на сфере
($dr=0$), либо на прямой ($d\Omega=0$). Это обусловлено тем, что
исходный лагранжиан задан в четырёхмерном пространстве-времени,
также как и векторные бозоны, индуцирующие геометрию.

Таким образом в данной моделе предполагается, что размер
элементарной частицы -- это  гравитационный радиус сильных
взаимодействий. По аналогии с нерелятивисткой гравитацией Ньютона
можно предложить формулу сильного взаимодействия между частицами
массы $m_1$ и $m_2$, как модель сильной гравитации:
\begin{equation}\label{p15_14}
F=-\kappa_s\frac{m_1m_2}{r^2}
\end{equation}

Вспомним, что рассматривается элементарная частица, влиянием спина и
заряда  которой  на геометрию пренебрегается. Пусть частица состоит
из двух кварков, Тогда возникает сила кулоновского притяжения и
притяжение сильной гравитации. Таким образом гамильтониан системы
имеет вид
\begin{equation}\label{p15_15}
H=\frac{\vec p^2_1}{2m_1}+\frac{\vec p^2_2}{2m_2}-\frac{Z_1Z_2e^2}{|\vec r_1-\vec r_2|}-\kappa_s\frac{m_1m_2}{|\vec r_1-\vec r_2|}
\end{equation}

Перейдём в систему центра инерции: $\vec p_1+\vec p_2=0$ с
приведённой массой $\mu$ и  поместим начало координат в точку, где
$\vec r=\vec r_1-\vec r_2$:
\begin{equation}\label{p15_16}
H=\frac{\vec p^2}{2\mu}-\frac{Z_q^2e^2}r-\kappa_s\frac{m_c^2}r
\end{equation}

В единицах $erg\cdot cm$ в системе СГС для верхних кварков, то есть
для u, c, t с  электрическим зарядом $+\frac23e$, найдём кулоновский
вклад
\begin{equation}\label{p15_17}
(Z_ce)^2=(\frac23\cdot4.8\cdot10^{-10}SGS_q)^2=10^{-19}
\end{equation}
а для нижних кварков, то есть для d, s, b кварков с электрическим
зарядом $-\frac13$  этот же вклад равен $0.25\cdot10^{-19}$.

Наконец кулоновское притяжение между нижними и верхними кварками даёт $0.5\cdot10^{-19}$.

С другой стороны, для u кварка массой 5 Мэв сильная гравитация также даёт:
\begin{equation}\label{p15_18}
\kappa_sm_u^2=(0.003\cdot1.67\cdot10^{-24}g)^2\cdot 5.4\cdot10^{31}\frac{ cm^3}{g\cdot cek^2}\approx1.38\cdot10^{-21}
\end{equation}
но так как этот вклад зависит от массы, то для разных кварков он
сильно различный.  Составим таблицу для различных значений $\kappa_s
m_im_k$ кварков $u,d,s,c,b=i,k$ массами и 3 Мэв, 5 Мэв, 100 Мэв, 1.3
Гэв и 4.5 Гэв соответственно
$$\begin{array}{cccccc}
&u&d&s&c&b\\
u&1.38\cdot10^{-21}&2.25\cdot10^{-21}&4.5\cdot10^{-20}&5.85\cdot10^{-19}&2\cdot10^{-18}\\
d&&3.75\cdot10^{-21}&7.5\cdot10^{-20}&9.75\cdot10^{-19}&3.4\cdot10^{-18}\\
s&&&1.5\cdot10^{-18}&1.95\cdot10^{-17}&6.75\cdot10^{-17}\\
c&&&&2.5\cdot10^{-16}&8.8\cdot10^{-16}\\
b&&&&&3\cdot10^{-15}
\end{array}$$
(включение в таблицу t-кварка не имеет смысла ввиду его большой
массы по сравнению с массой  $W^-,W^+$-бозонов, связанное состояние
которых образует постоянную связи сильного взаимодействия
$\kappa_s$)

Видим, что в первом приближении для s, c и b кварков можно
пренебречь кулоновским потенциалом,  оставляя только
сильно-гравитационное притяжение частицы и античастицы, для u и d
кварков наоборот, кулоновское притяжение частицы и античастицы более
значительно. В случае мезонов со скрытой странностью кулоновский
потенциал сравним с сильно-гравитационным потенциалом.

Под гравитационным радиусом метрика нестационарная, при этом материя
распределена на прямой.  Вне прямой выполняется  уравнение Эйнштейна
без материи $R_{ik}=0$. Но на прямой плотность энергии и давления
ненулевые, поэтому правая часть уравнений Эйнштейна на прямой
отлична от нуля и метрика имеет вид
\begin{equation}\label{p15_19}
ds^2=c^2dT^2-a^2(t)dr^2
\end{equation}
с некоторым $a(t)$. В такой метрике в соответствии с законом Хаббла
две точки, находящиеся  на рассоянии $r$, приближаются друг к другу
со скоростью
\begin{equation}\label{p15_20}
v=\frac {a'}ar=Hr
\end{equation}
поэтому квадрат полной энергии частицы есть
\begin{equation}\label{p15_21}
E^2=p^2c^2+m^2c^4=(mv)^2c^2+m^2c^4=(mcHr)^2+m^2c^4
\end{equation}

Откуда для больших $H$ получим
\begin{equation}\label{p15_22}
E=mcr\sqrt{H^2+c^2/r^2}\approx mcHr+\frac{mc^2}{2H}\cdot\frac1r
\end{equation}

Следовательно для больших импульсов имеет место линейный закон
увеличения энергии частицы,  что, как известно, является
экспериментальным фактом.

Заметим, что указанный рост потенциала с расстоянием, нисколько не
отрицает конфаймента, обусловленного глюонными полями, скорее его
дополняет, но с другой стороны.

Заметим, что под гравитационным радиусом допустимо, чтобы $r$ было
бы значительно больше  размера элементарной частицы, так как для
внешнего наблюдателя имеет значение величина $l=a(t)r$, которая при
малых $a(t)$ даёт необходимое $l$. Тем самым согласуется
представление о том, что комптоновский размер лёгких кварков
значительно больше размера элементарной частицы. С другой стороны,
параметр расстояния $r$ -- это параметр под горизонтом, поэтому при
переходе к длинам вне горизонта со стороны наблюдателя необходимо
вводить линейный множитель $a(t)$, по этой причине зависимость между
энергией частица $E$ и потенциалом $V(r)$ более сложная, нежелив
представлении (\ref{p15_18}).

\section{Заключение}
Таким образом ОТО рассматривается как О-теория в классическом
приближении задачи многих  тел. С математической точки зрения ОТО --
это ковариантное определение дуального простраства над пространством
состояний частиц. Ситуация близка к тому, что имеется
фундаментальная локальная теория электромагнетизма Максвелла и её
классическое представление в виде законов электростатики, закона
Ома, Кихгофа и т. д.

Здесь показано, что с локальной точки зрения для сохранения закона
сохранения заряда  на планковских временах и масштабах за счёт
неассоциативнго взаимодействия бозонных векторных полей D, D*, C и
E образуется связанное состояние массивных заряженных D-бозонов, в
результате чего D-бозоны теряют массу, а их взаимодействие с другими
частицами сводится к потенциальному взаимодействию, имеющему
калибровочную форму неинерциальной системы координат.

Физические величины строятся на базе исходного пространства и ему
дуального. Одно из  представлений дуального пространства, изоморфное
любому другому, строится с помощью метрического тензора. Кроме того,
появляется дополнительный потенциальный член, полностью определяемый
геометрическим характером взаимодействия. Поэтому на расстояниях,
больших планковских, уравнения должны иметь ковариантный тензорный
вид, то есть вид, предлагаемый ОТО.

В данной работе рассматривается общая схема построения теории
сильных взаимодействий.  Известно, что модель с линейным потенциалом
(\ref{p15_22}) и кулоновским потенциалом (\ref{p15_16}) хорошо
описывает спектроскопию кваркониев \cite{UFN}, поэтому
представляется интересным дальнейшее исследование предложенной в
этой работе модели сильных взаимодействий. Заметим, что цветная
симметрия $SU(3)$ в данном подходе не исключается при
соответствующем подборе коэффициентов $\alpha_i$ и $\beta_k$ в
(\ref{p14_1}) и (\ref{p14_2}), однако исследование этого вопроса не
входит в предмет данной работы.

Таким образом  в данной работе показано, что на классическом уровне
можно предложить модель сильных взаимодействий, обладающую важной
особенностью линейно растущего потенциала и невозможностью
существования свободных кварков. Кроме того, модель предлагает вид
сильных взаимодействий как сильной гравитации под горизонтом
Шварцшильда.

\section{Благодарности}
Работа  выполненена при частичной финансовой поддержки министерства
образования и науки Российской Федерации.

\end{document}